\documentclass[journal]{vgtc}                
\ifpdf
  \pdfoutput=1\relax                   
  \usepackage{graphicx}                
  \DeclareGraphicsExtensions{.pdf,.png,.jpg,.jpeg} 
\else
  \usepackage{graphicx}                
  \DeclareGraphicsExtensions{.eps}     
\fi%

\graphicspath{{figures/}{pictures/}{images/}{./}} 
\usepackage{amsmath}
\usepackage{amssymb}
\usepackage{microtype}                 
\PassOptionsToPackage{warn}{textcomp}  
\usepackage{textcomp}                  
\usepackage{mathptmx}                  
\usepackage{times}                     
\usepackage{cite}
\usepackage{comment}
\usepackage{tabu}                      
\usepackage{booktabs}                  
\usepackage{algorithm} 
\usepackage{algpseudocode}
\algdef{SE}{Begin}{End}{\textbf{begin}}{\textbf{end}}

\onlineid{0}

\vgtccategory{Research}
\vgtcpapertype{Journal}

\title{Visualizing Ensemble Predictions of Music Mood}

\author{Zelin Ye and Min Chen, \textit{Member, IEEE}}
\authorfooter{
\item
 Zelin Ye and Min Chen are with University of Oxford.\\
 E-mail: zelin.ye@linacre.ox.ac.uk, min.chen@oerc.ox.ac.uk.
}

\shortauthortitle{Biv \MakeLowercase{\textit{et al.}}: Global Illumination for Fun and Profit}

\abstract{Music mood classification has been a challenging problem in comparison with other music classification problems (e.g., genre, composer, or period). One solution for addressing this challenge is to use an ensemble of machine learning models. In this paper, we show that visualization techniques can effectively convey the popular prediction as well as uncertainty at different music sections along the temporal axis while enabling the analysis of individual ML models in conjunction with their application to different musical data. In addition to the traditional visual designs, such as stacked line graph, ThemeRiver, and pixel-based visualization, we introduce a new variant of ThemeRiver, called ``dual-flux ThemeRiver'', which allows viewers to observe and measure the most popular prediction more easily than stacked line graph and ThemeRiver. Together with pixel-based visualization, dual-flux ThemeRiver plots can also assist in model-development workflows, in addition to annotating music using ensemble model predictions.%
} 


\teaser{
  \centering
  \includegraphics[width=160mm]{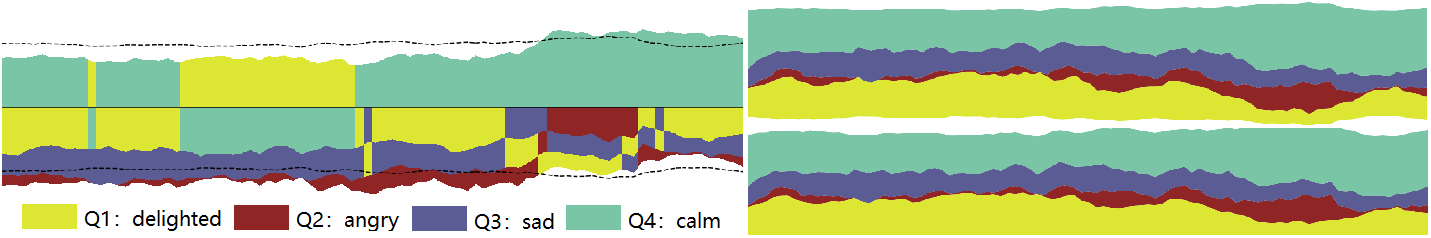}
  \caption{An ensemble of 210 machine learning (ML) models was used to predict music mood. The music shown is a segment from Partita V, Clavierubung part I, Bach BWV 829. From the original ThemeRiver (top right) and stacked line graph (bottom right), it is not always easy to see the ordering of the moods according to the votes, nor the points where the mood changes. A new visual design, \emph{dual-flux ThemeRiver} (left), enables more effective observation of such information. The upper flux shows the dominant mood, and the lower flux depicts the other moods in the descending order of received votes. In this figure, votes by ML models are weighted by the squared accuracy of ML models. The weighted 50\% lines depict a critical threshold over time.}
  \label{fig:teaser}
}

\vgtcinsertpkg

\begin{document}



\maketitle

\section{Introduction}
\label{sec:Introduction}

Music classification is an important family of techniques for enabling music categorization, recognition, segmentation, and retrieval.
In comparison with typical classification problems (e.g., genre and composer \cite{meng2005improving, kong2020large}), mood classification appears to be more challenging \cite{hu2017framework,bressan2017decision} without using lyrics features, as some of the music has relatively vague or complicated moods, while the genre can be clearly defined as using certain chords, instruments or rhythm pattern. The problem of music mood classification entails three challenges:

(a) Mood labels are typically assigned to a long piece of music rather than its parts because segmenting music paradoxically would need correct mood labelling for individual parts. 
    Extrapolating ``global'' ground truth labels to individual parts will inevitably lead to erroneous parts labels. For many music pieces, there is a flow of mood change across the music, and machine learning (ML) workflows have difficulties (i) to train a part-based model using ``global'' ground truth labels and (ii) to convey the uncertainty in mood classification effectively.

(b) The mood of music is inherently uncertain as a similar sequence of music scores may convey different moods. 
    Sometimes the same or similar notes can impress the listeners with different moods, e.g., by introducing an extra chord, playing the same notes in a different key, or changing the speed of a few notes.
    Some music pieces have more recognizable mood, e.g., \emph{delightful}, \emph{sad}, \emph{angry}, and \emph{calm}, while many other pieces may tingle between two moods or convey a mixture of moods as mood judgement can be subjective \cite{lu2005automatic, wieczorkowska2006multi}.
    In many ways, such uncertainty is also common in other mood classification problems, such as blog posts \cite{mishne2005experiments} or lyric text mining \cite{laurier2008multimodal, hu2009lyric}.%

(c) The ensemble ML models can potentially address challenges (a) and (b) partly by allowing different models to offer their predictions as ``opinions'' or ``votes''.
However, it is common for ensemble ML techniques to infer a single prediction from different opinions or votes, e.g., using weighted aggregation or a meta-model \cite{dietterich2000ensemble, polikar2006ensemble, sagi2018ensemble}.
    Such over-abstraction creates difficulties in observing how different parts of the music are voted by the ensemble of ML models or how individual model has performed for different music patterns.%

The above three challenges suggest that the ML technology can benefit from the assistance of visualization and visual analytics (VIS) techniques. To address challenge (a), VIS can reveal the performance of ML models in predicting part-labels after being trained with only ``global'' labels. To address (b), VIS can convey multiple moods that may potentially be present in individual parts of a music piece as well as the relative levels of their presence. To address (c), VIS can provide ensemble models that deliver multi-predictions with more complex visual representations for the model outputs. While VIS solutions for (a, b, c) can support ML experts, a VIS solution for (b) can reach out to a broader audience who are interested in the music mood.

In this work, we propose to visualize the multiple predictions by the ensemble of ML models, offering visual information that the traditional statistical aggregation cannot offer.
In particular, we considered three visual designs, stacked line graph and ThemeRiver \cite{havre2000themeriver, havre2002themeriver} for aggregated opinions and pixel-based visualization for constituent predictions of individual models.
During our effort for using such visual designs to support our ensemble ML workflows, we discovered a variant of ThemeRiver, which is more effective than the original ThemeRiver and stacked line graph.
Fig. \ref{fig:teaser} shows an example of such a variant, and a section of the music is also shown in a stacked line graph and an original ThemeRiver plot.
This variant allows viewers to observe the change in the opinions of ML models more easily, and perceive whether the majority opinion has passed a critical threshold more precisely. We refer to this new variant as \emph{dual-flux ThemeRiver}.

While we use the dual-flux ThemeRiver variant to address the need for temporal and collective mood visualization in challenges (a) and (b), we use pixel-based visualization to address the need for visualizing the performance of many individual models in the above challenge (c). The combination of dual-flux ThemeRiver and pixel-based visualization also enables the juxtaposition of the overall mood flow and the contribution of individual models.

The overall contribution of this work is the introduction of appropriate visualization techniques into ensemble ML workflows. In particular, we firstly introduce \emph{dual-flux ThemeRiver} (Section \ref{sec:ThemeRiver}) as a novel visual representation to enable effective visual communication of multiple mood predictions by many ML models. This is crucial to address the above challenges (a) and (b) and to support pixel-based visualization in terms of challenge (c). Secondly, we provide ML model developers with investigative visualization for evaluating the performance of many models in a context sensitive to individual parts of the music (Section \ref{sec:PixelVis}).
This allows ML developers to observe a huge amount of information that statistical indicators cannot convey, facilitating the transformation from
actionable information to ML-developmental decisions (e.g., weighting schemes in ensemble ML). Hence, this addresses challenge (c) that is routinely encountered in developing an ensemble ML technique involving many models.

We present two case studies (Sections \ref{sec:Op57} and \ref{sec:Validation}) to demonstrate the usability of the proposed techniques. We conduct an analytical evaluation to reason the shortcomings of the existing solutions and the advantages and potential limitations of the proposed techniques (Section \ref{sec:AnalyticalEvaluation}). We report the evaluation by music and ML experts, confirming the merits of this work (Section \ref{sec:ExpertEvaluation}).

\section{Related Work}
\label{sec:RelatedWork}
This paper relates to visualization and visual analytics (VIS), music mood classification and machine learning (ML) in several aspects. In this section, we give a brief overview of each aspect.


\vspace{1mm}
\noindent\textbf{Time-series Visualization.}
The survey in the book \cite{aigner2011survey} offers a comprehensive list of visual designs up to 2011.
Recently Fang et al. provided an updated survey on the topic \cite{fang2020survey}.
There are many versions of line graphs for visualizing time series.
Harris \cite{harris1999information} proposed a visual design using area graph stacking lines on top of each other as a variant of the original layer area graph. 
%
Among variants of line graphs, ThemeRiver is a popular visual design and is highly relevant to this work.
Harve et al. \cite{havre2000themeriver,havre2002themeriver} proposed the ThemeRiver design for visualizing textual data or large document collections.
There are many variants of ThemeRiver, including
the spiral variant by Jiang et al. \cite{jiang2016spiral},
a multi-dimensional variant by Kalouli et al. \cite{kalouli2019parhistvis},
a variant with trend line and change indicators by Shi et al. \cite{shi2012rankexplorer},
variants with labelling and wiggling deformation by Byron and Wattenberg \cite{byron2008stacked},
a variant with heatmap by Hashimoto and Matsushita \cite{hashimoto2012heat}.
With all these designs, viewers have to determine the ordering of different layers by visually estimating the thickness of each layer at each time step.

To depict ordering change explicitly, Javed et al. \cite{javed2010graphical} introduced an overlapped design, where multiple layers are superimposed and ordered from lower values (at front) to higher values (at back).
However, while the change of ordering can be perceived easily, the visual estimation of the thickness of each layer at each step requires cognitive reasoning about the perceived color-filled shapes as well as the occluded shapes.

\vspace{1mm}
\noindent\textbf{Ensemble Data Visualization.}
Ensemble data visualization is a collection of visual designs and techniques for visualizing data derived by ensemble modelling and simulation.
Wang et al. provided a survey on VIS techniques for ensemble data \cite{wang2018visualization}.
A variety of VIS techniques have been developed, including
a collection of interactive visualization techniques by 
Potter et al. \cite{potter2009ensemble};
topology preserving mappings by Baruque et al. \cite{baruque2009ensemble}
an approach called Noodles for visualizing uncertainty of ensemble weather modelling by Sanyal et al. \cite{sanyal2010noodles},
%
%

a cluster-based ensemble visualization technique by
Kumpf et al. \cite{kumpf2017visualizing},
several visual representations under a Lagrangian framework by
Hummel et al. \cite{hummel2013comparative},
contour boxplots for uncertainty visualization by
Whitaker et al. \cite{whitaker2013contour},
3D animation of integrated ensemble data by
Phadke et al. \cite{phadke2012exploring},
a multi-view tool with continuous parallel coordinate plots and ensemble bars by
Chen et al. \cite{chen2015uncertainty},
a multi-view tool with geospatial visualization, spaghetti plots, cluster-based parallel coordinates by 
Jarema et al. \cite{jarema2016comparative},
and a multi-view tool with plots for displaying time-hierarchical clustering, stacked spatiotemporal data, time-step animation, and space-time surface by Ferstl et al. \cite{ferstl2016time}.


Because almost all these works deal with complex outputs (e.g., 3D geometry and spatiotemporal data), much focus was placed on displaying and exploring the output data resulting from ensemble simulation. Our work differs from these works in two aspects. (i) While the predictions in each time series are simple nominal values, the temporal dimension contains much more semantically-meaningful information. (ii) There is a stronger need to observe individual model's predictions at individual time steps and there are 210 such models.
For these two reasons, we focused on the ThemeRiver and pixel-based visualization.


\vspace{1mm}
\noindent\textbf{Visualization for Machine Learning.}
Sacha et al. \cite{sacha2018vis4ml} provided an ontology for showing various places in ML workflows which can benefit from visualization, followed by Tam et al. \cite{tam2016analysis}, who used information theory to estimate the benefit of visualization in ML workflows. We are going to apply a similar strategy for ML.
Rauber et al. \cite{rauber2016visualizing} proposed an overview visual design for comparing model structures, in which we find the ensemble models for time-series data in music should also have similar functions and to compare within the model.
In the VIS4ML ontology outlined by Sacha et al. \cite{sacha2018vis4ml}, this work is part of the step of ``Evaluate-Model'' in ML workflows, including the process blocks ``Model-Testing'', ``Quality-Analysis'', and ``Understanding-Model''. Our contributions include a new general-purpose visual design as well as new application-specific solutions and experiences.

\vspace{1mm}
\noindent\textbf{Ensemble Machine Learning Models.}
%
Ensemble ML techniques have been developed to address the problem that a single ML model may not be able to reach the required accuracy.
Dietterich et al.\cite{dietterich2000ensemble}, Polikar et al. \cite{polikar2006ensemble}, and Sagi et al. \cite{sagi2018ensemble}  provided surveys on ensemble ML techniques.
These surveys showed that the majority of ensemble ML models were designed to improve model prediction by combining different opinions offered by models in the ensemble. The aggregated decision is commonly obtained using a weighting function or a machine-learned meta-model.
Opitz and MaClin \cite{opitz1999popular} studied how each classifier and dataset would affect the results of ensemble classification.
Rokach et al. \cite{rokach2010ensemble} discussed various ensemble classifiers techniques, including their weighting functions.
Many weighting functions have been proposed, including distribution summation by Clark et al. \cite{clark1991rule} and performance weighting by Opitz \cite{opitz1999feature}.
For meta-learning, ``stacking'' is a popular approach developed by Wolpert et al. \cite{wolpert1992stacked}, and ``mixture of experts'' is another popular method proposed by Jacobs et al. \cite{jacobs1991adaptive}.
Omari and Figueiras-Vidal \cite{omari2015post} proposed a post-aggregation method with combined weighting and meta-learning.


In the ensemble ML literature, some researchers reported the applications of multi-label classification, e.g., semantic annotation \cite{qi2007correlative} and music mood classification \cite{wieczorkowska2006multi}.
There are mainly two approaches for multi-label classification \cite{tsoumakas2007multi}: algorithm adaptation and problem transformation.
With the  algorithm adaptation approach, multi-label predictions are generated directly by multi-label classification models, e.g., kNN \cite{zhang2007ml}, adaboost \cite{schapire2000boostexter} and neural network \cite{zhang2006multilabel}.
With the problem transformation approach, a multi-label problem is decomposed into multiple single-label problems \cite{boutell2004learning}. 
The Binary Relevance (BR) method considers each label as an independent class and trains a single-label model independently \cite{madjarov2012extensive}.
The Label Powerset (LP) method considers different a new combination of labels as a new class \cite{tsoumakas2009mining}).
The RA$k$EL method reduces the complexity of LP by using an ensemble of multiple LP classifiers \cite{tsoumakas2010random}.
The Classifier Chain (CC) method adapts BR by chaining a set of independent models in a particular order \cite{read2011classifier}.

In this work, we use ensemble ML for multi-label classification and we adopted the problem transformation approach.

\vspace{1mm}
\noindent\textbf{Machine Learning for Music and Music Mood Classification.}
Basili et al. \cite{basili2004classification} proposed several ML methods for music classification and explored the impact of different music features on the model accuracy.
In the literature, ML techniques used for music classification include:
support vector machine by Xu et al. \cite{xu2005automatic,liu2009cultural},
adaboost by Bergstra et al. \cite{bergstra2006aggregate},
and transformer-based methods by Zhuang et al. \cite{zhuang2020music, zhao2021musicoder}.
Choi et al. \cite{choi2015topic} proposed to use topic modeling techniques for categorizing users' interpretation of song lyrics for music classification.

Laurier et al. \cite{laurier2007audio} used support vector machine for music mood classification.
Hu et al. \cite{hu2009lyric} showed that lyrics features could be useful for music mood classification.
In the field of music, it is widely known that mood classification can be subjective and experts may have different opinions about the same piece of music \cite{lu2005automatic, wieczorkowska2006multi}.
Hence, the labels in the training data are treated as the closest labels rather than the ground truth.
There are mainly two approaches for mood modelling: categorical and dimensional \cite{kim2011music}. Hevner \cite{hevner1935affective, hevner1937affective} studied the categorical approach and clustered 67 moods into 8 groups.
Because the adjectives for mood labelling could be ambiguous and misleading, the alternative models by Rusell \cite{russell1980circumplex} and Thayer \cite{thayer1990biopsychology} are more widely used.
Both models have been used in ML-based mood classification. For example,
Jim et al. \cite{kim2011music} used Thayer's model, while a few others used Rusell's model \cite{skowronek2006ground, hu2010lyrics, hu2017framework}. Our work followed Rusell's model and used the 4-mood model dataset \cite{panda2018musical, panda2018novel} associated with this model.%

\section{Ensemble Music Mood Classification}
\label{sec:EnsembleModels}
Music classification is to determine if a piece of music falls into a certain category in relation to a concept. Example concepts are genre, composer, period, mood, and instrument.
Many ML techniques have been applied to music classification, including 
support vector machine \cite{dhanalakshmi2009classification},
convolutional neural networks \cite{lidy2016parallel},
recurrent neural networks \cite{jakubik2017evaluation}, and
transformer \cite{zhuang2020music}.
Some concepts, such as genre, composer, and period, are associated with a whole piece of music. For these classification problems, it is relatively easy to obtain accurate labels for training and testing ML models. Many recent techniques have already achieved very good results. Our own experiments showed that it is not too challenging to attain 90\% accuracy for such problems.

However, the mood classification is not so easy as a ``global'' label for a piece of music is not necessary to be applicable to every constituent part. Music is a narrative, where moods may change over time. Meanwhile, it is difficult to label individual parts of music, as it requires accurate segmentation of parts according to mood. Hence, paradoxically, mood classification and segmentation are two mutually dependent problems. This work is concerned with training ML models for mood classification without ideal mood labels. The ML models are required to determine music mood at individual parts, but can only learn to do so using the less accurate ``global'' labels.

Our overall approach is to train many models using different ML techniques and training settings.
From the perspectives of ML and the two challenges, (a) and (b), in Section \ref{sec:Introduction}, different ML techniques and training settings can enable models to learn different temporal features from ``global'' labels so that these models may judge individual parts differently.
An ensemble of model predictions resembles multiple opinions in the human judgement of music mood.
In the following subsections, we describe the training data, the ML methods used, the ensemble strategy, and the problems encountered in our ML workflow.

%
\vspace{2mm}
\noindent\textbf{Training and Testing Dataset.}
For training our models, we used the public domain dataset, \emph{4Q Audio Emotion Dataset} \cite{panda2018musical,panda2018novel}.
It includes about 900 pieces of clips, with each associated with a mood label from \textit{delighted}, \textit{angry}, \textit{sad}, and \textit{calm} according to Rusell`s model \cite{russell1980circumplex}. Because the ML models are required to determine music mood at individual parts, we defined the smallest part as one second. For the music clips in the dataset, we used the slide-window mechanism to transform each clip into parts of $i \in [1, 30]$ seconds. For example, for a 30s clip, there will be one part of 30s, two parts of 29s, three parts of 28s, and so on.
The ``global'' label of the 30s clip is extrapolated to all parts.
Our first premise is that if a ``global'' label of a clip indicates a mood \emph{Q}, it normally applies to most parts of the clip. Hence, probabilistically, mood \emph{Q} is mostly correct among these parts.
Our second premise is that learning from inaccurate labels is common in many ML applications, such as topic modelling \cite{choi2015topic} and text mood classification \cite{hu2009lyric}.

To apply and evaluate our ML models, we also used another public domain dataset, \emph{Multimodal Sheet Music Dataset} \cite{dorfer2018learning}. These music clips are longer and are accompanied by music scores, but do not have mood labels.
The music in Fig. \ref{fig:teaser} is from this dataset.

\vspace{2mm}
\noindent\textbf{Feature-based Machine Learning.}
In this work, we use one of our ML workflows as an example to demonstrate the need for data visualization. This workflow focuses on feature-based ML using techniques belonging to the decision tree (DT) family.
The basic DT algorithms include CART\cite{breiman2017classification} and C4.5 \cite{quinlan2014c4}.
More sophisticated techniques in the family include
bagging \cite{breiman1996bagging},
adaboost \cite{freund1997decision},
gradient boosting decision tree (GBDT)\cite{friedman2001greedy},
XGBoost \cite{chen2015xgboost},
random forest (RT) \cite{liaw2002classification},
and Deep Forest (gcForest) \cite{zhou2019deep}.
These techniques use different strategies to improve the basic DT. We are particularly interested in how their performance is affected by training with music parts of different interval lengths.

Feature extraction is an integral part of all algorithms and techniques in the DT family. We extracted 55 features ($f_1, \ldots, f_{55}$) from spectrogram from all music parts used for training and testing, including:

\begin{itemize}
    \vspace{-2.5mm}
    \item \textbf{chroma\_stft}: $f_1, f_2$. The normalized energy.
    \vspace{-2.5mm}
    \item \textbf{rms}: $f_3, f_4$. The root-mean-square values.
    \vspace{-2.5mm}
    \item \textbf{spectral\_centroid}: $f_5, f_6$. The mean of the normalized magnitude of the spectrogram (a distribution over frequency bins).
    \vspace{-2.5mm}
    \item \textbf{spectral\_bandwidth}: $f_7, f_8$. The frequency bandwidth.
    \vspace{-2.5mm}
    \item \textbf{spectral\_rolloff}: $f_9, f_{10}$. The centre frequency where 85\% of the energy of the spectrum is contained in this bin and the bins below.%
    \vspace{-2.5mm}
    \item \textbf{zero\_crossing\_rate}: $f_{11}, f_{12}$. The zero-crossing rate of the data.%
    \vspace{-2.5mm}
    \item \textbf{tempogram}: $f_{13}, f_{14}$. Local auto-correlation of the onset strength envelope.
    \vspace{-2.5mm}
    \item \textbf{MFCCs}: $f_{15} - f_{54}$. Decompose the waveform into different frequency bands. The first 20 MFCCs.
    \vspace{-2.5mm}
    \item \textbf{tempo}: $f_{55}$. Estimation of beats per minute.
\end{itemize}
For each feature extracted (except the tempo), we compute the mean and variance for feature integration.

The common wisdom in the area of ML for music \cite{srinivasan2004harmonicity,zhang2004audio} is that features of a shorter temporal window are suitable for conveying information about timbre, and those of a longer temporal window are suitable for conveying information of modulation, beat, mood, vocal, and so on.
However, there has not been a thorough investigation of the size of the window.
Some previous work used window sizes from 10ms to 10s \cite{foote2001beat}. It was found that 10ms segments were too short to have any information for classification \cite{meng2005improving}. The uncertainty about window sizes motivated us to explore as many reasonable options as possible and to train ML models with window sizes of 1 to 30 seconds.

\vspace{2mm}
\noindent\textbf{Ensemble Music Models.}
In our ML workflow, 
we train ML models using seven different methods in the DT family. They are basic DT (CART), bagging, adaboost, gradient boosting decision tree (GBDT), XGBoost, random forest (RF), and Deep Forest (gcForest). 
With each method, we train 30 ML models for features extracted from different interval lengths (i.e., $i \in [1, 30]$).
Our ML workflow thus generates 210 models for mood classification.
Our initial goals for ensemble ML were (i) to identify the best model from the 210 models and (ii) to use these models to compute aggregated ensemble predictions.

\begin{figure}[t]
    \centering
    \includegraphics[width=80mm,height=35mm]{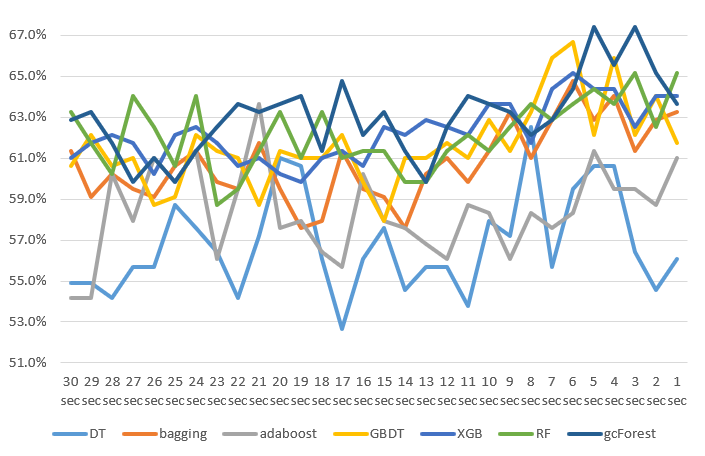}\\[-2mm]
    \caption{Accuracy of the 210 ML models trained in this work.}
    \label{fig:ModelAccuracy}
    \vspace{-4mm}
\end{figure}

As shown in Fig. \ref{fig:ModelAccuracy},
the basic DT has the lowest accuracy with an average accuracy of 56.8\% for ML models corresponding to the 30 interval lengths.
The next is adaboost, with an average accuracy of 58.4\%.
This is followed by bagging trees (60.8\%), GBDT (61.6\%), Random Forest (62.2\%), and XGBoost (62.3\%).
The best is gcForest with an average accuracy of 63.1\%.
In particular, the joint best individual models are
gcForest operating at 3s and 5s intervals respectively and both achieving 67.4\% overall accuracy. The second best is GBDT at 5s interval (66.7\%), and the third-best is GBDT at 4s (65.9\%).

We tried many typical ensemble strategies, including the maximum number of (a) unweighted votes, and (b) votes weighted some quality measures of individual models, including  ($b_1$) by the F1-score, ($b_2$) by the overall accuracy, ($b_3$) by the squared overall accuracy, ($b_4$) by mood-specific class-accuracy (based on confusion matrices), and ($b_5$) by squared mood-specific class-accuracy. 
The performance of these ensemble models was disappointing.
We found that these statistical measures did not help ML model developers to think, and especially those with music knowledge could not use their knowledge in analyzing the model performance in relation to individual music clips at different temporal locations.
As discussed in Section \ref{sec:Introduction}, one major reason is that the accuracy measures themselves are not reliable because the given global labels for 30s intervals are not the ground truth for the individual music clips of 1s to 29s intervals.
%

\vspace{2mm}
\noindent\textbf{Requirements for Data Visualization.}
According to Chen and Ebert's methodology for improving a workflow, when statistic measures abstract information too quickly, data visualization can offer a remedy \cite{chen2019ontological}.
Having realized that the statistical measures could not enable the two initial goals mentioned earlier, the ML developer working on music mood prediction refocused the development effort on VIS. While the ML developer had an intuition about several VIS requirements, a relatively full set of requirements, as listed below, were identified iteratively during the development life-cycle, including the analytical evaluation and expert evaluation reported in Section \ref{sec:Results}.
We consider two groups of target users, music experts and ML developers.


\begin{itemize}
    \vspace{-2mm}
    \item R$_1$. Both would like to observe how ensemble models collectively voted on individual sections of music, so someone with music knowledge can reason if the voting results are sensible or not.
    \vspace{-2mm}
    \item R$_2$. Both would like to observe how ensemble models are collectively influenced by the less accurate ``global'' ground truth labels in parts of the music where the mood changes.
    \vspace{-2mm}
    \item R$_3$. Both would like to locate where ensemble models voted for a mood change so we can relate such changes with the corresponding music score.
    \vspace{-2mm}
    \item R$_4$. Both would like to see the dominant opinion of ML models, the second dominant opinion, the third, and so on, and music experts would like to exercise their own interpretations of the different predictions generated by an ensemble of ML models.
    \vspace{-2mm}
    \item R$_5$. Both would like ideally to identify visual representations that can be used to accompany music for non-experts.
    \vspace{-2mm}
    \item R$_6$. ML developers would like to observe sub-groups of models (e.g., by methods and interval length) to compare their performance with the ensemble group.
    \vspace{-2mm}
    \item R$_7$. ML developers would like to observe individual models' performance to compare their performance with the ensemble group and related subgroups.
\end{itemize}

\begin{figure*}[t]
    \begin{tabular}{@{}c@{\hspace{4mm}}c@{\hspace{1mm}}c@{\hspace{4mm}}c@{\hspace{1mm}}c@{\hspace{4mm}}c@{\hspace{1mm}}c@{}}
        &
        \multicolumn{2}{c}{\small{\textsf{(a,b,c,d) unweighted level of mood}}} &
        \multicolumn{2}{c}{\small{\textsf{(e,f,g) weighted level of mood}}} &
        \multicolumn{2}{c}{\small{\textsf{(h,i,j) weighted \& normalised level of mood}}}\\[1mm]
        \includegraphics[height=13mm]{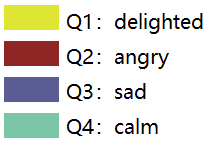} & \includegraphics[height=13mm]{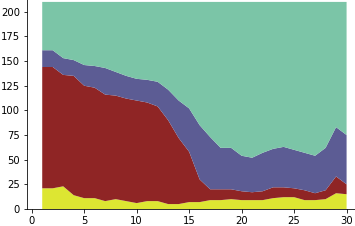} & \includegraphics[height=13mm]{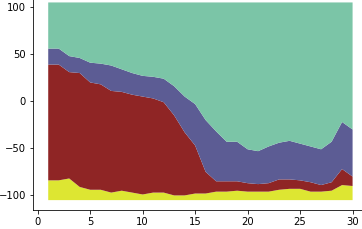} & \includegraphics[height=13mm]{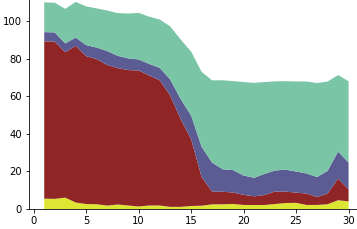} & \includegraphics[height=13mm]{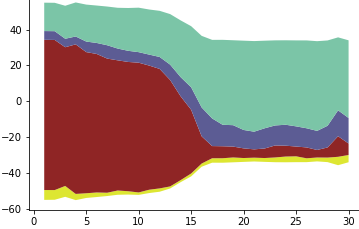} & \includegraphics[height=13mm]{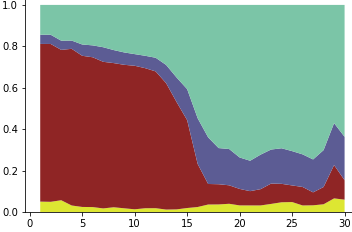} & \includegraphics[height=13mm]{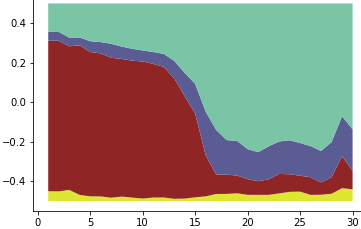}\\
         & \small{(a) stacked line graph} & \small{(b) ThemeRiver} & \small{(e) stacked line graph} & \small{(f) ThemeRiver} & \small{(h) stacked line graph} & \small{(i) ThemeRiver}\\[1mm]
    \end{tabular}
    \centering
    \begin{tabular}{@{}c@{\hspace{4mm}}c@{\hspace{4mm}}c@{\hspace{4mm}}c@{}}
        \includegraphics[height=26mm]{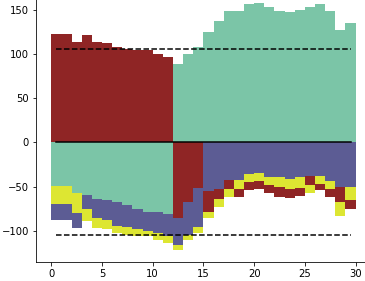} & \includegraphics[height=26mm]{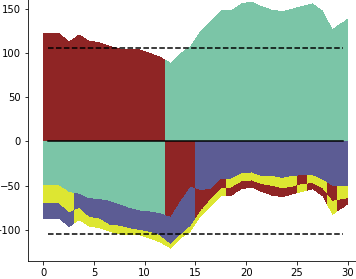} & \includegraphics[height=26mm]{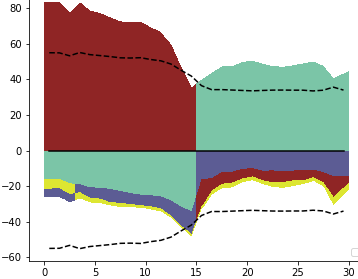} & \includegraphics[height=26mm]{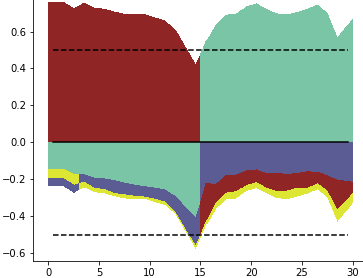}\\
        \small{(c) basic dual-flux ThemeRiver} & \small{(d) smoothed dual-flux ThemeRiver} & \small{(g) smoothed dual-flux ThemeRiver} & \small{(j) smoothed dual-flux ThemeRiver}\\[-1mm]
        \small{with measurable main stem} & \small{with measurable main stem} & \small{with measurable main stem} & \small{with measurable main stem}
    \end{tabular}
    \caption{Three visual designs for depicting time-varying ensemble predictions, namely stacked line graph (a), original ThemeRiver (b), and dual-flux ThemeRiver (c,d). In (a,b,c,d), the width of each ``prediction'' stem is the unweighted sum of the number of models that made that prediction. The river width is thus the total number of models (i.e., 210 in this case).
    In (e,f,g) the width of each stem is the weighted sum by the squared class-accuracy of those models that made the prediction. In (h,i,j), the weights are normalized, i.e., the total number of models for each second is normalized to 1.}
    \label{fig:VisualDesigns}
    \vspace{-4mm}
\end{figure*}

\section{Visualizing Ensemble Model Predictions}
\label{sec:Visualization}
As mentioned in Section \ref{sec:EnsembleModels}, our ML workflow results in 210 ML models for mood classification. Given a piece of music, these models will scan sections of the music and offer their predictions section-by-section. While the accuracy analysis in Section \ref{sec:EnsembleModels} may be indicative as to the overall performance of a model, they have the following shortcomings:

(1) The term ``accuracy'' is in fact rather misleading. As mentioned earlier, each mood label available to the ML workflow is for the whole piece of music rather than individual sections.
    When a piece of music is labelled as ``sad'', it may be safe to consider most parts are ``sad'' but rather unsafe to conclude every section is ``sad''.
    Therefore, when a ML model fails to assign a ``sad'' prediction to a section of the music concerned, it does not necessarily mean that the ML model made an error.

(2) The ``accuracy'' measure (or any other commonly-used quality measure, e.g., F1-score) is thus uncertain. The accuracy measure cannot convey the ``opinions'' of different models for each section of music. Some of the model predictions might be considered as errors in calculating the ``accuracy'' measure, but are actually contributing meaningful ``opinions'' in judging the mood of individual sections, for which the ground truth is unknown.

(3) The ``accuracy'' measure is not detailed enough for ML model-developers to investigate the behaviour of each model in relation to different music patterns.

One major underlying cause of (1) and (2) is that uncertainty is inherently part of music and other forms of arts. It may be better for us to embrace the uncertainty by conveying different ``opinions'' instead of an aggregated prediction. 
Another major cause that underlies (2) and (3) is that statistical measures compress information too quickly \cite{chen2019ontological}, and visualization can alleviate such shortcomings by slowing down the pace of information loss.

In this section, we examine the use of two families of visualization techniques, \emph{stacked line graph} and \emph{pixel-based visualization} for bringing back some lost information.

\subsection{Stacked line graph, ThemeRiver, Dual-flux ThemeRiver}
\label{sec:ThemeRiver}
As discussed in Section \ref{sec:EnsembleModels}, different ML models process a piece of music in different section lengths. 
We define the \emph{unit-section length} as the maximal common denominator of all section lengths. For our ML models, it happens to be 1 second.
Given a piece of music $\mathcal{M}$, we can first divide it into $n$ unit-sections, $\mathcal{M} = \{s_1, s_2, \ldots, s_n \}$,
and then translate the predictions of each model to the predictions for these unit-sections. Given $m$ ML models, we have $m$ model-specific time series (with categorical values) as $T_i = \{c_{i,1}, c_{i,2}, \ldots, c_{i,n} \},\; i=1,2,\ldots,m$.
%

%
For each time interval $s_j$, we can compute the total number of models that predicted a particular mood, choosing from ``delighted'', ``angry'', ``sad'', and ``calm''. This results in four summary time series (with numerical values for ensemble classification results).
\begin{equation} \label{eq:CTS}
\begin{aligned}
    T_\text{delighted} =& \{v_{d,1}, v_{d,2}, \ldots, v_{d,n} \} \quad
    T_\text{angry} = \{v_{a,1}, v_{a,2}, \ldots, v_{a,n} \}\\
    T_\text{sad} =& \{v_{s,1}, v_{s,2}, \ldots, v_{s,n} \} \hspace{5mm}
    T_\text{calm} = \{v_{c,1}, v_{c,2}, \ldots, v_{c,n} \}\\
\end{aligned}
\end{equation}

The time series in the Eq. \ref{eq:CTS} are naturally suitable for the stacked line graph and ThemeRiver as shown in Figs. \ref{fig:VisualDesigns}(a,b). With these two visual designs, it is not easy for one to observe which mood is the most popular opinion, especially at the temporal locations where two or more moods received similar numbers of votes from ML models.

We thus introduce a new visual design, which is referred to as \emph{dual-flux ThemeRiver}, the basic version of which displays each time step as several stacked bars.
As shown in Fig. \ref{fig:VisualDesigns}(c), we divide the ``river'' into two main fluxes. The upper flux always displays the data of the mood receiving the most popular votes.
The lower flux displays the data of other moods, ordered by the number of votes received from ML models. We can also depict critical thresholds using grid-lines in both fluxes, allowing viewers to perceive whether the most popular opinion has passed the critical threshold. For example in Fig. \ref{fig:VisualDesigns}(c), two grid-lines indicate the 50\% threshold of 210 votes. 
Note that it is not easy to introduce such grid-lines in the original ThemeRiver representation (e.g., in Fig. \ref{fig:VisualDesigns}(e)).
As demonstrated in Figs. \ref{fig:VisualDesigns}(c,d,g,j), with the dual-flux ThemeRiver, one can observe the change in the popularity of the opinions and whether they pass the threshold line more easily than the stacked line graph and the original ThemeRiver. To achieve the same smoothed effect as the stacked line graph and ThemeRiver in Figs. \ref{fig:VisualDesigns}(a,b), we also developed an algorithm (see Supplemental Materials) that can handle any potential change of ordering between time steps as shown in Fig. \ref{fig:VisualDesigns}(d), which improves the visual design in Fig. \ref{fig:VisualDesigns}(c).


In ensemble modelling, it is common to assign different weights to votes of models according to a quality measure, e.g., accuracy values, F1-scores, or the squared version \cite{large2019probabilistic}. Once such weights are introduced, the width of the river (in $y$ direction) may vary temporally (in $x$-direction).
In this work, we consider that the accuracy of each model depends on what decision it is making.
Hence we obtain the \emph{class-accuracy} of each model based on its confusion matrix. 
Fig. \ref{fig:VisualDesigns}(e,f,g) shows three examples of ensemble predictions for the three visual designs, where each vote is weighted by the squared class-accuracy of the model.
One may observe that the delighted mood (yellow) at the left-bottom of (c,d) has shrunk and moved downwards in Fig. \ref{fig:VisualDesigns}(g). This is because the delighted mood in that music segment was largely returned by models that are less accurate in predicting delight.
Often weighting by class-accuracy can reduce the frequency of order switching caused by ``voting noise''. 
One may also notice that for the dual-flux ThemeRiver in Fig. \ref{fig:VisualDesigns}(g), the critical threshold is defined as 50\% of the sum of all weighted votes, and it also varies temporally.


Comparing Figs. \ref{fig:VisualDesigns}(a,b,c,d) and Figs. \ref{fig:VisualDesigns}(e,f,g), we notice the change of the popular opinion in the middle of the plot (i.e., 13$\sim$15th seconds), where the most popular \emph{angry} opinion (red) was replaced by the \emph{calm} opinion (green).
Because the ``global'' label for the whole excerpt is \emph{angry}, we consider that the weighted aggregation is more meaningful than the unweighted one.
A close look at the sample shows that the first half has a lot of strong drum beats and sound, with low frequency at a fast speed without melody. This ends up being classified as angry. The second half has synthesized sound only without a meaningful melody.
The \emph{calm} opinion is not necessarily incorrect.
Many ML models must have made such a prediction based on their learning from other music pieces.
This also explains why \emph{angry} opinion becomes the second least popular opinion in the second half of the music.

Naturally, we may consider to normalize the weighted votes such that the total number of votes will remain uniform temporally. Figs. \ref{fig:VisualDesigns}(h,i,j) show the normalized versions of (e,f,g) respectively. Generally normalization makes the stacked line graph and the original ThemeRiver into rectangular boxes, while the dual-flux ThemeRiver still displays a fair amount of dynamics. While the normalized version may appear to some as more intuitive or less complicated, it may be mistaken as ``one model, one vote'' as with Fig. \ref{fig:VisualDesigns}(d). Meanwhile, the unnormalized version in Fig. \ref{fig:VisualDesigns}(g) not only indicates the presence of weighting but also conveys some uncertainty information. For example, the second section in Fig. \ref{fig:VisualDesigns}(g) is narrower than the first section, which indicates the mood classification is less certain because some highly-weighted models are minority voters. In the rest of the paper, we will focus on the smoothed, weighted, and unnormalized version.

\begin{figure*}[ht]
    \centering
    \begin{tabular}{@{\hspace{4mm}}c@{\hspace{30mm}}c@{\hspace{30mm}}c@{\hspace{12mm}}}
        \includegraphics[width=25mm,height=12mm]{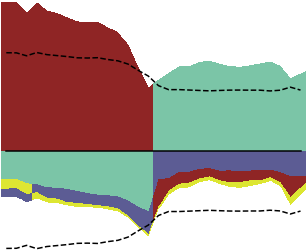} &
        \includegraphics[width=25mm,height=12mm]{Figures/5/fig5.png} &
        \includegraphics[width=25mm,height=12mm]{Figures/5/fig5.png}\\
    \end{tabular}
    \begin{tabular}{@{}r@{\hspace{2mm}}c@{\hspace{2mm}}l@{\hspace{2mm}}%
                       r@{\hspace{2mm}}c@{\hspace{2mm}}l@{\hspace{8mm}}%
                       r@{\hspace{2mm}}c@{\hspace{2mm}}l@{}}
        \includegraphics[height=125mm]{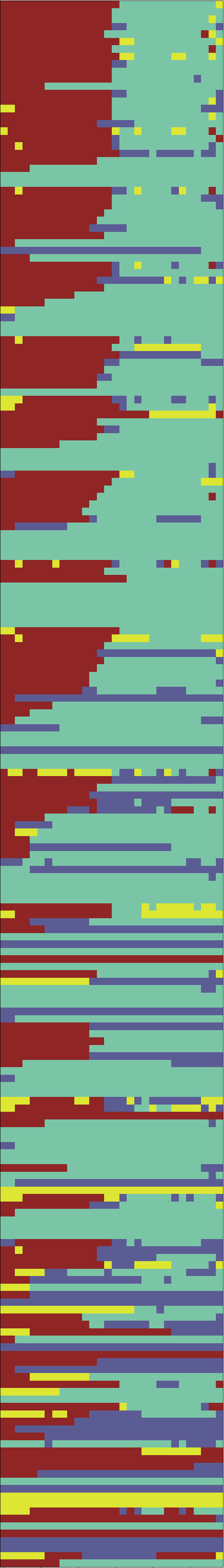} &
        \includegraphics[height=125mm]{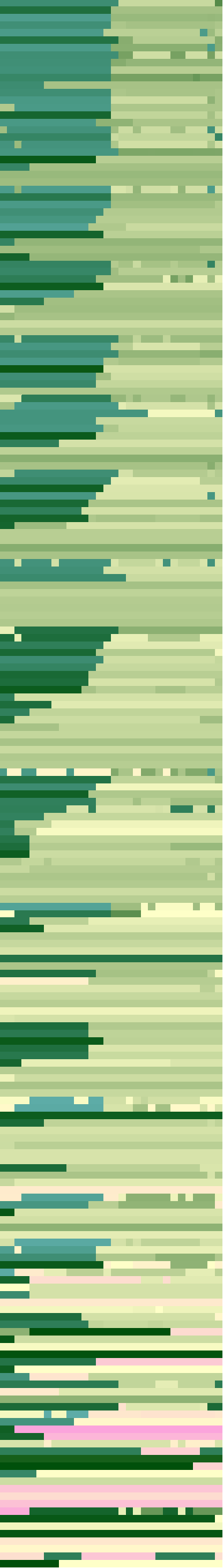} & 
        \includegraphics[width=4mm,height=125mm]{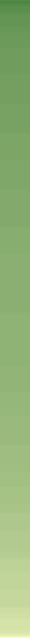} &
        \includegraphics[height=125mm]{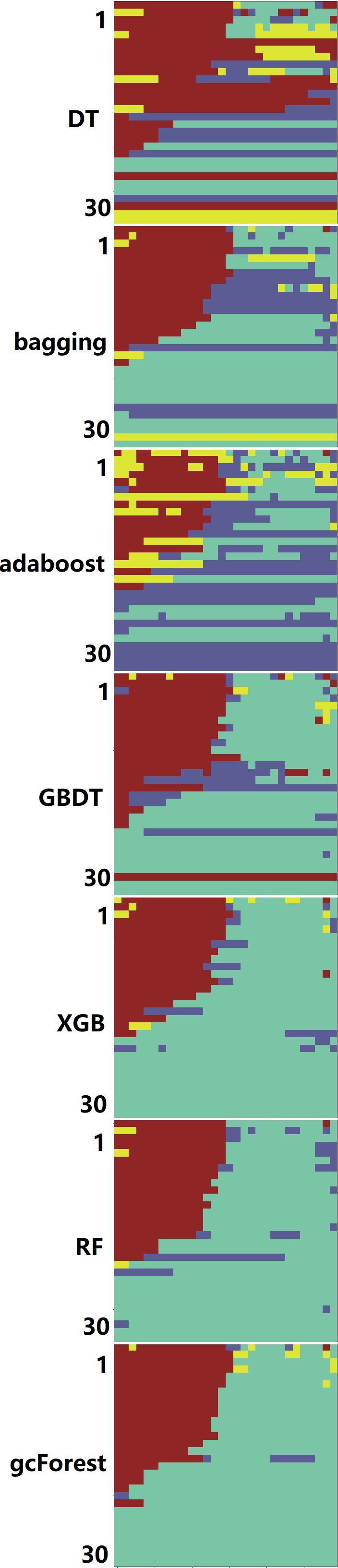} &
        \includegraphics[height=125mm]{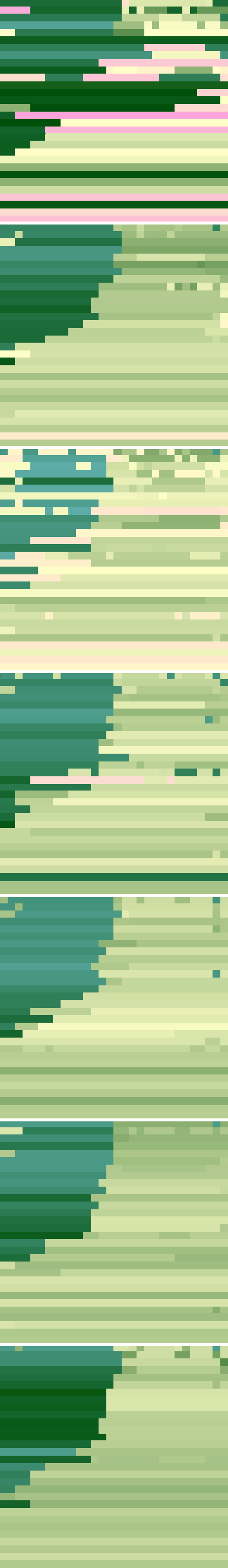} & 
        \includegraphics[width=4mm,height=125mm]{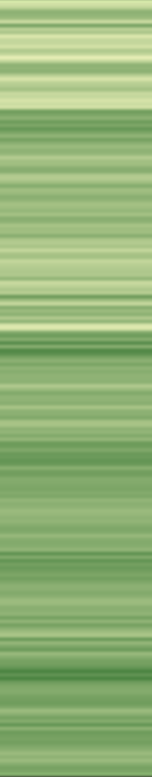} &
        \includegraphics[height=125mm]{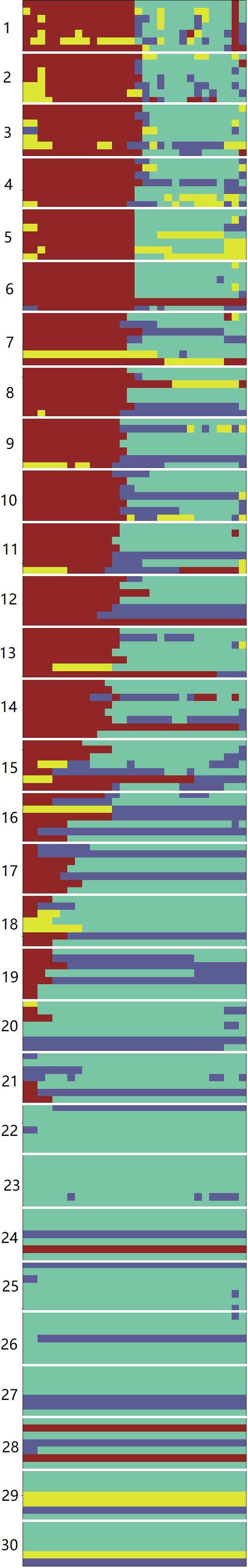} &
        \includegraphics[height=125mm]{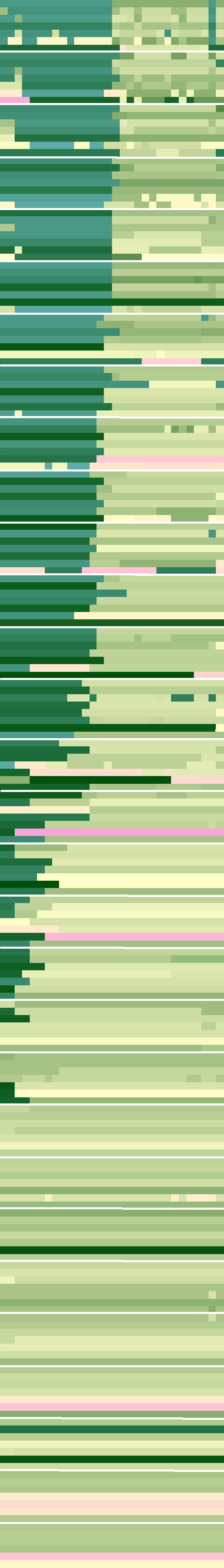} & 
        \includegraphics[width=4mm,height=125mm]{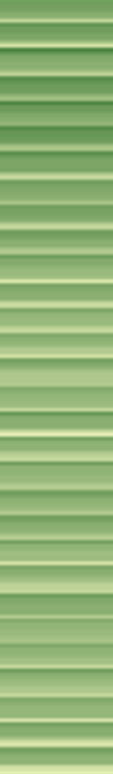}\\
        \multicolumn{3}{c}{(a) sorted by accuracy} &
        \multicolumn{3}{c}{(b) by methods then interval length} &
        \multicolumn{3}{c}{(c) by interval length then accuracy} \\[2mm]
    \end{tabular}\\
    \centering
    \includegraphics[height=3mm]{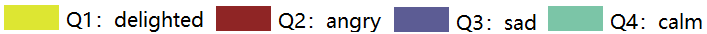} \quad
    \includegraphics[height=4mm]{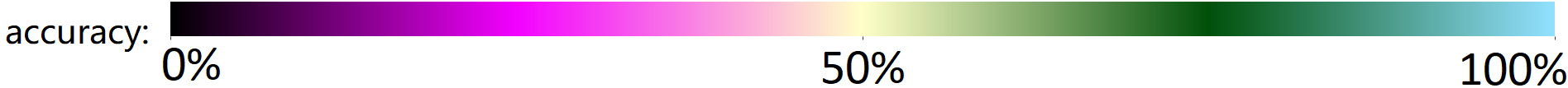}
    \caption{Pixel-based visualization of the predictions by individual models. Each model corresponds to a row of pixels, while models are sorted along the $y$-axis using different sorting schemes. Color-mapped model accuracy is shown on the right of each plot. The same weighted dual-flux ThemeRiver (as Fig. \ref{fig:VisualDesigns}(g)) is shown above each pixel-based plot as a reference to how class-accuracy may affect the dual-flux ThemeRiver.}
    \label{fig:PixelBased}
    \vspace{-4mm}
\end{figure*}

\subsection{Pixel-based Visualization with Sorted/Clustered Models}
\label{sec:PixelVis}
In an ensemble modelling workflow, inevitably, we are interested in observing individual models' performance as well as comparing groups of models with similar attributes. Pixel-based visualization can provide an effective overview for observing the performance of many models in the context of different music.
For example, as discussed in Section \ref{sec:EnsembleModels}, the models in the ensemble were trained with different ML techniques, or sound features used by models are of different time intervals.


Fig. \ref{fig:PixelBased} shows three sets of pixel-based visualization, all corresponding to the weighted ensemble predictions in Fig. \ref{fig:VisualDesigns}(g).
In each set of pixel-based visualization, the first pixel-map shows the prediction results.
In the pixel-map, each pixel is a model's prediction for a time step of the music concerned (1 second in our case), and each row depicts a model's predictions per unit-section (30 seconds in our case). Each pixel-map essentially depicts the data structure of $T_i = \{c_{i,1}, c_{i,2}, \ldots, c_{i,n} \}$, except that the rows are sorted differently to enable the comparison of models with different grouping strategies.

The second pixel-map shows the class-accuracy of each prediction.
For example, the top-left and bottom-left pixels in the first pixel-map indicate two angry predictions, while the corresponding pixels in the second pixel-map indicate two very different levels of class-accuracy. When we produce the weighted dual-flux ThemeRiver on the top of these two pixel-maps, the two angry predictions are weighted differently and contribute differently to the width of the first time step of the river.

Based on the discussion in the previous subsection, we consider that the popular predictions of \emph{angry} (red) in the first section and the predictions of \emph{calm} (green) in the second section are all correct.

In Fig. \ref{fig:PixelBased}(a), models (i.e., rows) are sorted according to accuracy. The accuracy values of models are depicted using a separate column on the right of the two pixel-maps.
We can observe that most models in the upper half of the plot (i.e., more accurate models) can separate \emph{angry} and \emph{calm} well.
A good number of models predicted the whole section as \emph{calm}. 
We consider that their predictions in the second half are correct, while their votes in the first half are questionable since the `global'' ground truth label is \emph{angry}.
In the lower part of the plot (less accurate models), the predictions seem to be much less consistent.
However, a few models that predicted the whole section as \emph{angry} are all located in the lower half of the plot, suggesting that these are ``accidental'' correlations with the `global'' ground truth label.

In Fig. \ref{fig:PixelBased}(b), models are sorted by the ML methods and then interval length (from 1 to 30). 
We can observe that gcForest, RF, and XGB methods offer opinions consistent with the popular views depicted in the upper flux of the dual-flux ThemeRiver (top), while adaboost is the least consistent, followed by DT and bagging. The more accurate models of GBDT are consistent, but not with the less accurate ones.

Using Fig. \ref{fig:PixelBased}(b), we can also observe the impact of interval length.
The shape of the red pattern is similar across all seven ML methods, indicating that models with short interval lengths are more accurate. For the first half of the music, interestingly, the pixel block for the bagging method shows a pattern similar to gcForst, RF, XGB, and GBDT, but has more \emph{sad} opinions (blue) in the middle (interval lengths 7$\sim$12), suggesting that the bagging method may likely confuse \emph{calm} with \emph{sad} if they are trained with certain interval lengths.
Meanwhile, DT and adaboost also made many \emph{sad} predictions.
However, they appear less coherent than bagging when they are sorted by interval length in (c). Since models produced by these two ML methods are least accurate in the traditional sense (Section \ref{sec:EnsembleModels}), pixel-based visualization confirms that they are generally not as good as the other 5 ML methods by having the average accuracy of 56.8\% and 58.4\% while others are all above 60\%.

\begin{figure*}[t]
    \centering
    \includegraphics[height=18mm,width=145mm]{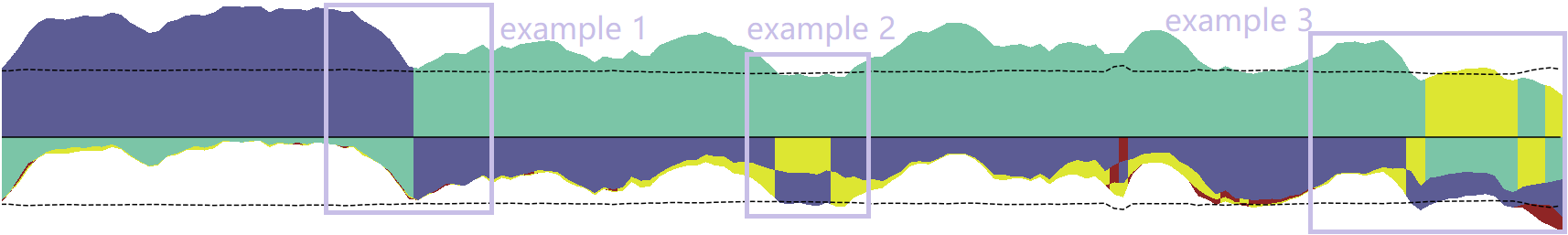}\\
    \begin{tabular}{@{}c@{\hspace{8mm}}c@{\hspace{8mm}}c@{}}
        \includegraphics[height=16mm,width=50mm]{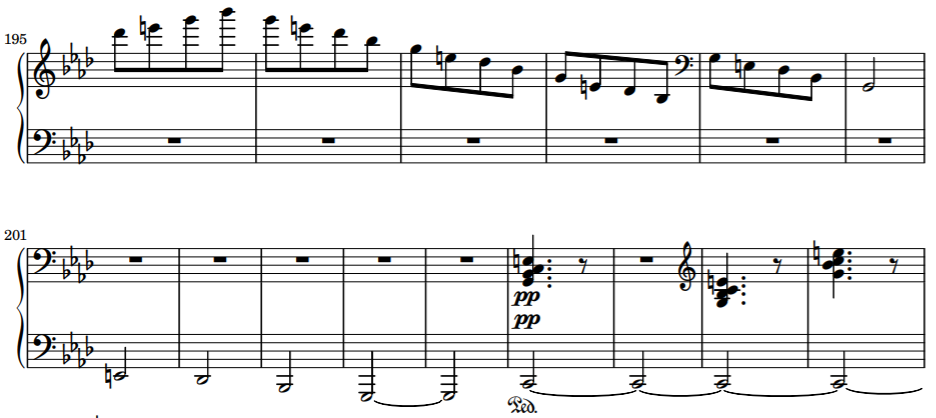} &
        \includegraphics[height=16mm,width=50mm]{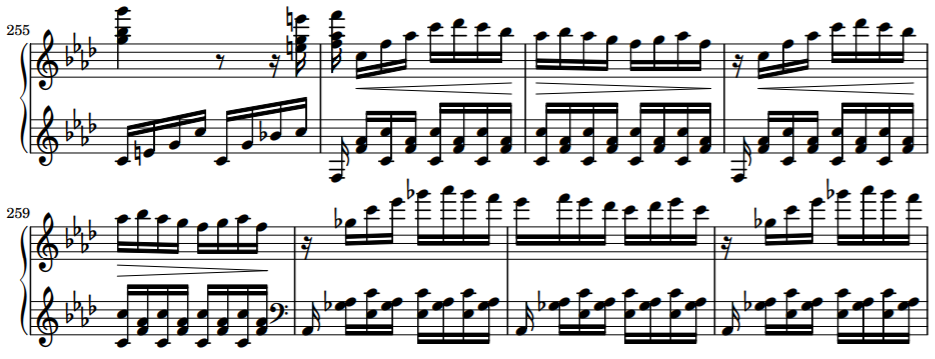} &
        \includegraphics[height=16mm,width=50mm]{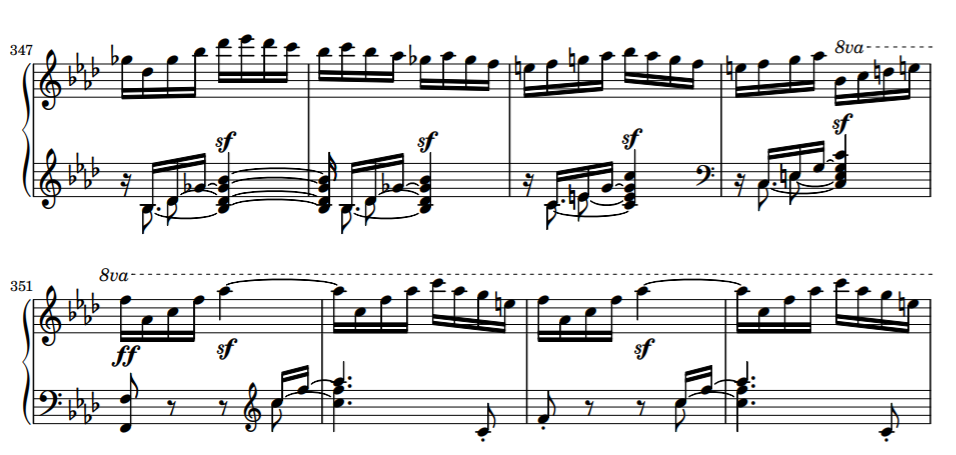} \\[-1mm]
        \small{bar 195 $\sim$ 209 (example 1)} &
        \small{bar 255 $\sim$ 262 (example 2)} &
        \small{bar 347 $\sim$ 354 (example 3)} \\
        \includegraphics[height=16mm,width=50mm]{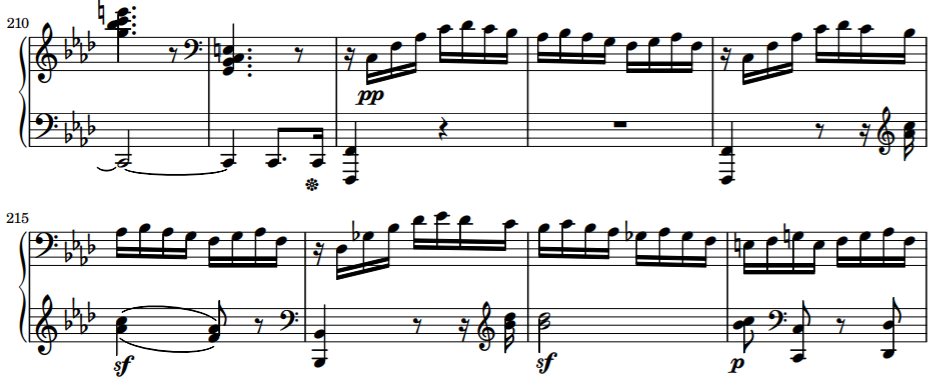} &
        \includegraphics[height=16mm,width=50mm]{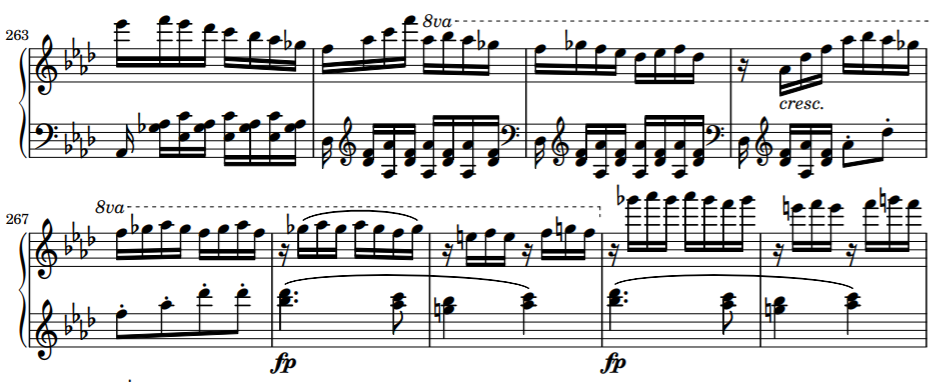} &
        \includegraphics[height=16mm,width=50mm]{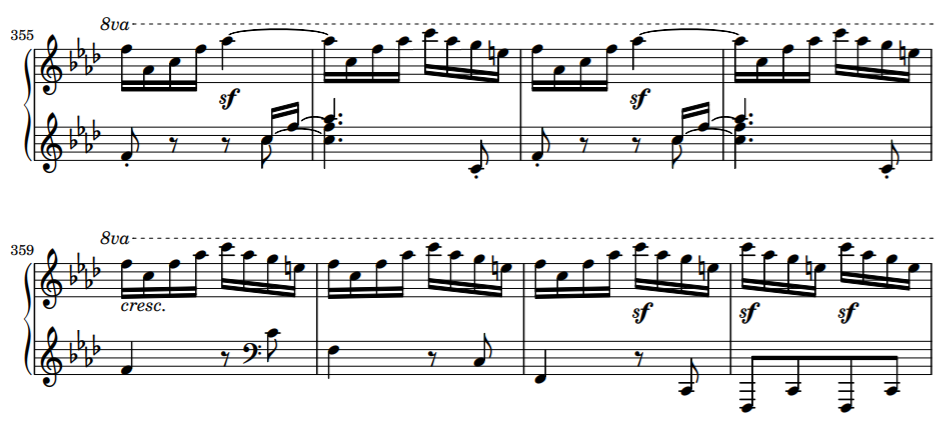} \\[-1mm]
        \small{bar 210 $\sim$ 218 (example 1)} &
        \small{bar 263 $\sim$ 271 (example 2)} &
        \small{bar 355 $\sim$ 362 (example 3)} \\
    \end{tabular}
    \caption{Example 1 to 3: dual-flux ThemeRiver for Piano Sonata Op 57 ($3^{rd}$ Movement).}
    \label{fig:Op57}
    \vspace{-4mm}
\end{figure*}

Fig. \ref{fig:PixelBased}(c) further confirms the impact of interval length.
This plot also allows us to make a critical judgement about training ML models with relatively long interval lengths. On the surface, if an interval length is closer to the length used for ``global'' ground truth labelling, the model evaluation appears to be more ``trustworthy''. However, we consider the ``global'' ground truth labels are not necessarily correct for every ``local'' part of the music piece concerned and our goal is to obtain models that can collectively make more correct predictions for the ``local'' parts.
Hence the pixel-based visualization in Fig. \ref{fig:PixelBased} would suggest trusting more the models trained with short interval lengths.
Of course, such a suggestion needs to be validated by the testing data for other music examples. Pixel-based visualization allowed ML model developers to perform such validation quickly.  


In Fig. \ref{fig:PixelBased}(c), models are sorted by the time interval length first, then the accuracy among the 7 ML models for each time interval. Here we can observe that the result of (c) is perfectly matched with the previous analysis from (b). Models with 22, 23, 25, 26, 27, 29, and 30 second time intervals have absolutely no red color. Only a few long segments are giving the angry mood, while most of the red color comes from the shorter segments, and there is a decreasing trend as the time interval length increases. This result also matches the dual-flux ThemeRiver above the pixel-based visualization, as the dominant mood started with angry, and was replaced by calm. In terms of the model accuracy, there is no huge difference among models of different time intervals, and the only slightly darker section is around 4 to 6 seconds.

\section{Case Studies and Evaluation}
\label{sec:Results}
In this section, we first describe two case studies, one focuses on the use of dual-flux ThemeRiver, and another focuses on the combined use of pixel-based visualization and dual-flux ThemeRiver. We then provide an analytical evaluation of the visual designs and the VIS4ML workflow as well as a human-centered evaluation involving six independent experts from ML and music disciplines.



\subsection{Case Study: Piano Sonata Op 57} 
\label{sec:Op57}
Fig. \ref{fig:Op57} depicts the ensemble predictions of the music mood by 210 ML models for the $3^{rd}$ Movement in Piano Sonata Op 57 by Beethoven, which is also known as Appassionata (Italian: passionate).
This is one of Beethoven's most famous and technically difficult sonatas. Three sections of mood changes were highlighted, and the corresponding music scores are also shown in the figure.

In the first example, the upper flux shows that a segment receiving votes of \emph{sad} ends at bar 211, and the dominant votes change to \emph{calm}.
From the music theory, the segment of \emph{sad} is part of \emph{development}, and the segment of \emph{calm} is part of recapitulation that leads to the end.

Before bar 211, the music is still slow, consisting of bass notes and chords. When the recapitulation starts, it suddenly turns into a series of notes at a normal speed from bar 211. This is when the switching happens in Example 1. Here the ensemble models collectively and precisely detected the change of music content and the mood, from a slow, \emph{sad} segment into a segment played at a normal speed, which leads to the dominance of \emph{calm}. Since it is still played in F minor, there is a considerable amount of \emph{sad} mood, which is meaningfully shown as the second dominant mood in the lower flux.

The music in the second example in Fig. \ref{fig:Op57} is a connecting episode and played in D flat major, which could cause an increase of \emph{delighted}. The dominant votes in the upper flux have not changed, as the playing style is similar to the previous segment (bar 212 $\sim$ 255, in F minor). The only change is the key being transposed to a major key. Therefore, although the music does not change much, the change of the second dominant mood in the dual-flux ThemeRiver shows that it is different.

In the third example, as the music approaches the end as part of the Coda, there is a slight change before the finish. The first 3 bars from 347 still repeat the same pattern as in Example 2 on the right hand from bar 256 to 267.
From bar 350, it is suddenly played in a higher octave, which might cause the dominant mood to be \emph{delighted}. In Fig. \ref{fig:Op57}, we also find the \emph{delighted} mood does not appear from nowhere, but gradually becomes dominant in bar 350. This matches the score, which indicates a continuous sequence of notes played by the right hand.

This case study shows that the collective predictions of ensemble models are meaningful and consistent with the music score. The dual-flux ThemeRiver enables the identification of music pattern changes, which are conveyed in the sense of mood changes. The ability to see the first and second dominant opinions is the major merit of dual-flux ThemeRiver over the original ThemeRiver and stacked line graph.

\begin{figure*}[t]
    \raggedright
    \hspace{15mm}ex1\hspace{8.5mm}ex2\hspace{8.5mm}ex3\hspace{9mm}ex4%
    \hspace{7.5mm}ex5\hspace{8.5mm}ex6\hspace{9mm}ex7\hspace{8mm}ex8%
    \hspace{7.5mm}ex9\hspace{7.5mm}ex10\hspace{7mm}ex11\hspace{7mm}ex12\\
    \centering
    \begin{tabular}{@{}r@{\hspace{1mm}}l@{\hspace{1mm}}r@{\hspace{1mm}}l@{\hspace{1mm}}r@{\hspace{1mm}}l@{\hspace{1mm}}r@{\hspace{1mm}}l@{}@{}r@{\hspace{1mm}}l@{\hspace{1mm}}r@{\hspace{1mm}}l@{\hspace{1mm}}r@{\hspace{1mm}}l@{\hspace{1mm}}r@{\hspace{1mm}}l@{}@{}r@{\hspace{1mm}}l@{\hspace{1mm}}r@{\hspace{1mm}}l@{\hspace{1mm}}r@{\hspace{1mm}}l@{\hspace{1mm}}r@{\hspace{1mm}}l@{}}
        \includegraphics[width=10.5mm,height=6mm]{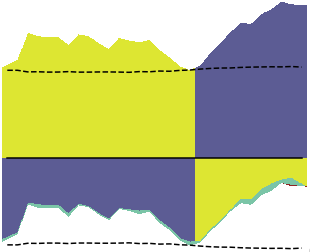} & &
        \includegraphics[width=10.5mm,height=6mm]{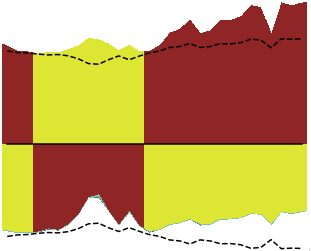} & &
        \includegraphics[width=10.5mm,height=6mm]{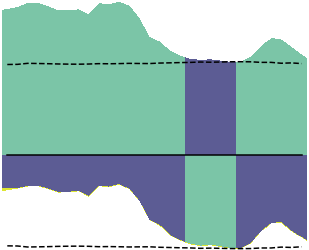} & &
        \includegraphics[width=10.5mm,height=6mm]{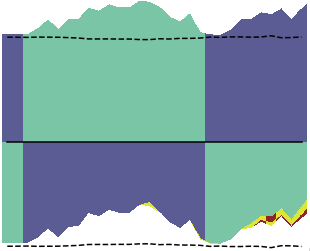} & &
        \includegraphics[width=10.5mm,height=6mm]{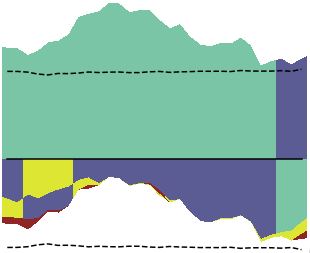} & &
        \includegraphics[width=10.5mm,height=6mm]{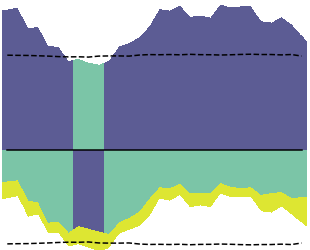} & &
        \includegraphics[width=10.5mm,height=6mm]{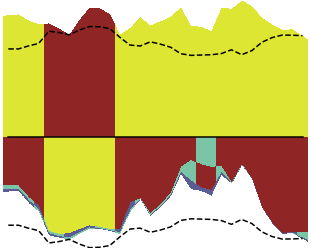} & &
        \includegraphics[width=10.5mm,height=6mm]{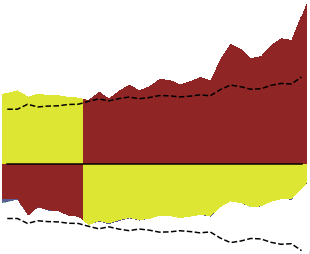} & &
        \includegraphics[width=10.5mm,height=6mm]{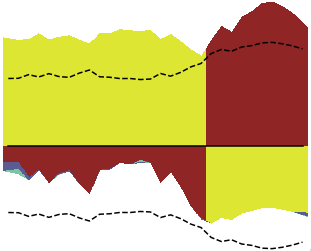} & &
        \includegraphics[width=10.5mm,height=6mm]{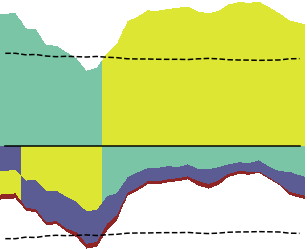} & &
        \includegraphics[width=10.5mm,height=6mm]{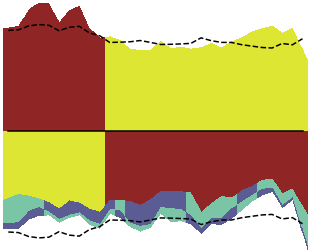} & &
        \includegraphics[width=10.5mm,height=6mm]{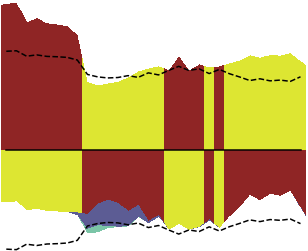}\\
        \includegraphics[height=68mm]{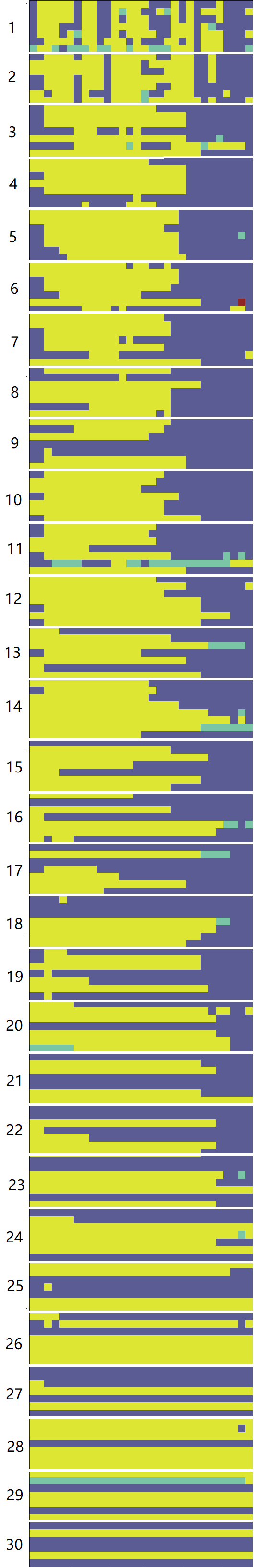} & &
        \includegraphics[height=68mm]{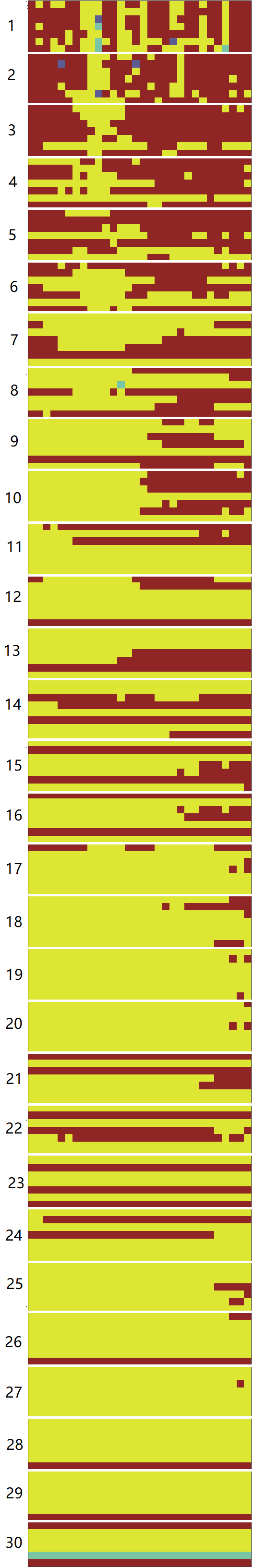} & &
        \includegraphics[height=68mm]{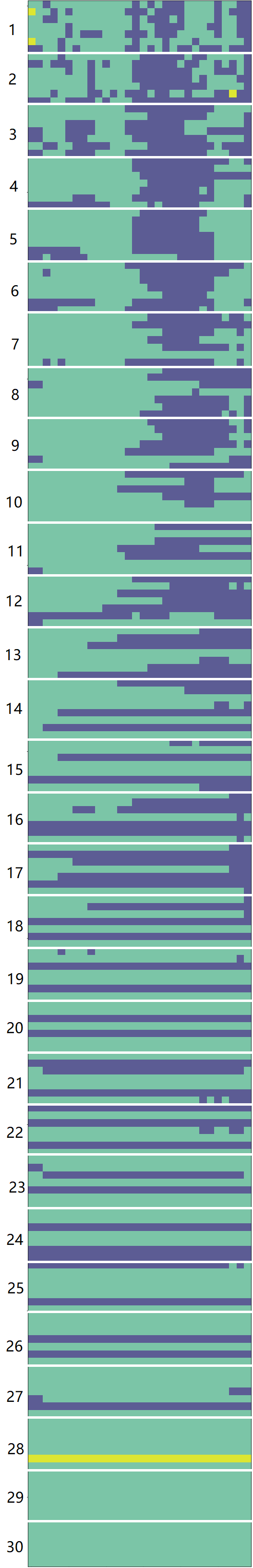} & &
        \includegraphics[height=68mm]{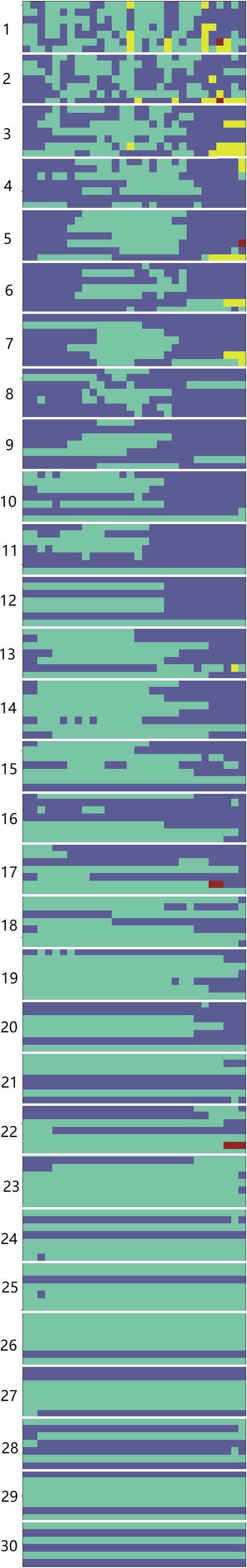} & &
        \includegraphics[height=68mm]{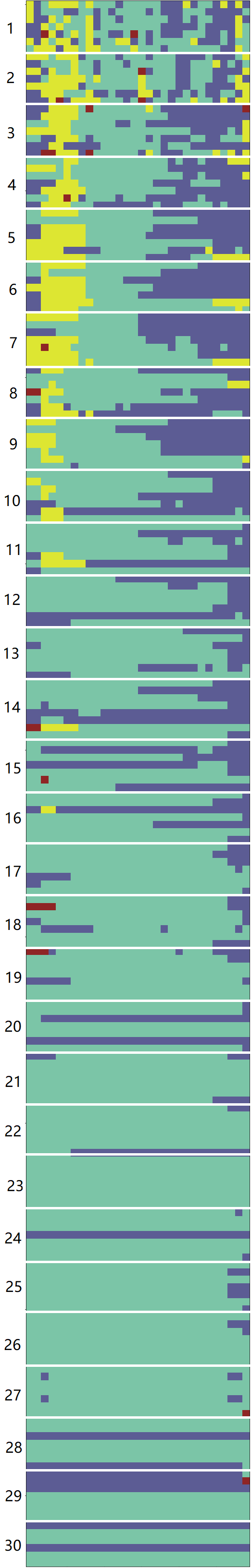} & &
        \includegraphics[height=68mm]{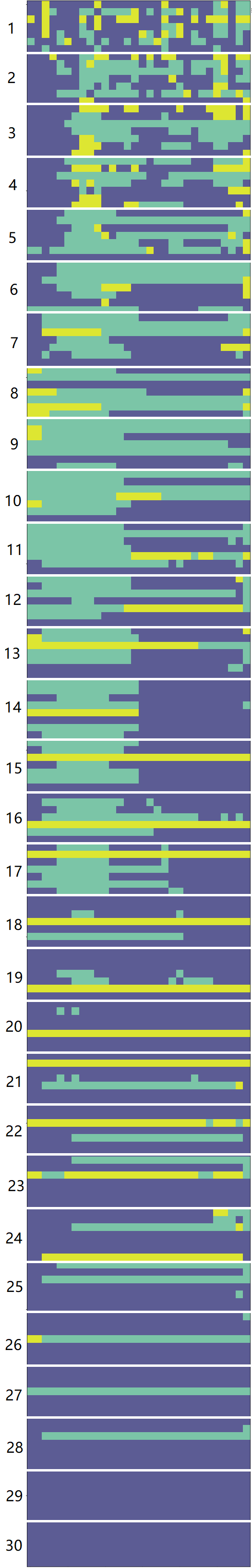} & &
        \includegraphics[height=68mm]{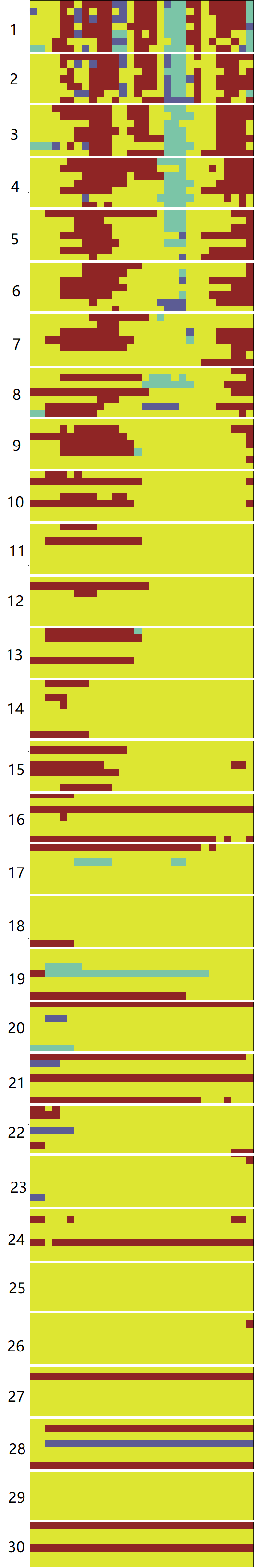} & &
        \includegraphics[height=68mm]{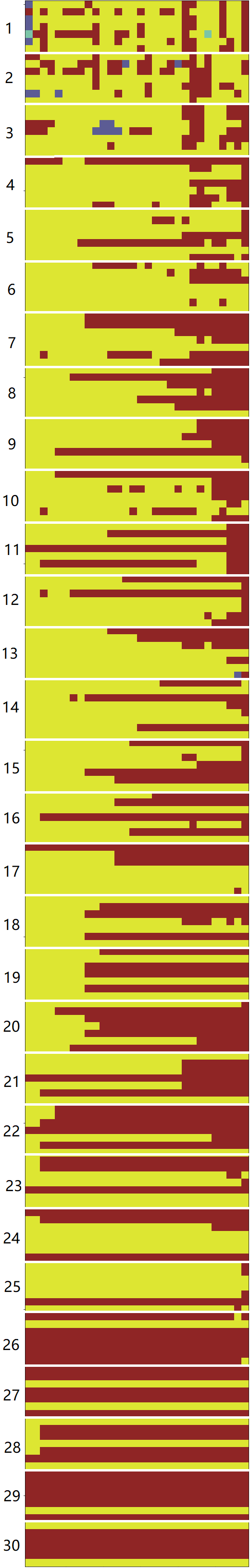} & &
        \includegraphics[height=68mm]{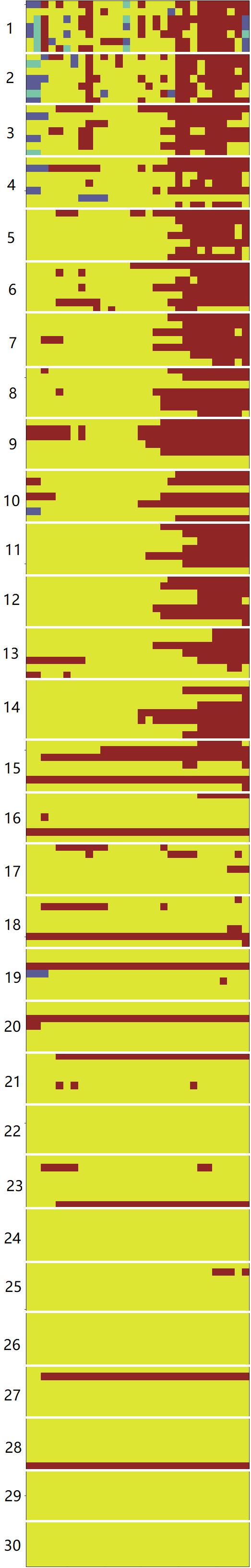} & &
        \includegraphics[height=68mm]{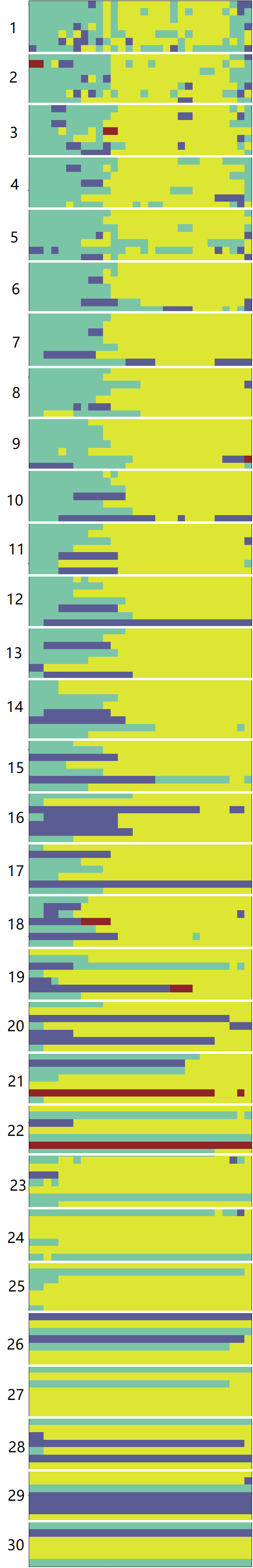} & &
        \includegraphics[height=68mm]{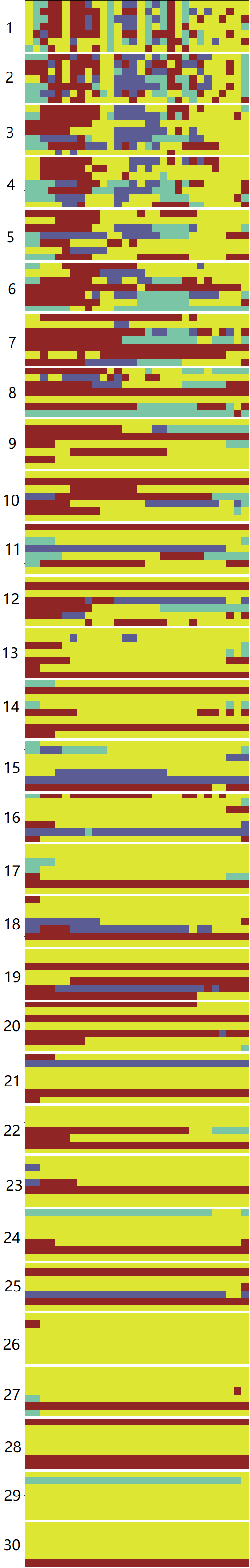} & &
        \includegraphics[height=68mm]{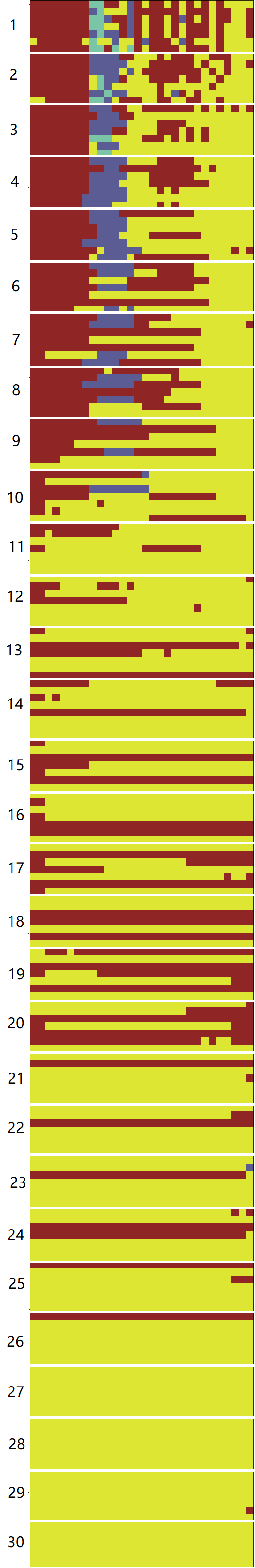} & \includegraphics[width=3.5mm,height=68mm]{Figures/section 5/p3.1.png}\\
    \end{tabular}\\
    \centering
    \includegraphics[height=3mm]{Figures/legend_row.png} \quad
    \includegraphics[height=3.5mm]{Figures/legend_color.png}
    \caption{Pixel-based visualization with the aid of dual-flux ThemeRiver, for evaluating a hypothesis that ML models trained with features of smaller intervals can deal with mood change better. The 12 dual-flux ThemeRiver show examples of mood changes and each corresponding pixel visualization shows the predictions of individual models, sorted by interval lengths. To evaluate the hypothesis, one can observe the correlation between the upper flux of the dual-flux ThemeRiver and different parts of the pixel-based visualization.}
    \label{fig:Validation}
    \vspace{-4mm}
\end{figure*}

\subsection{Case Study 2: Hypothesis Evaluation}
\label{sec:Validation}
Recall our discussion in Section \ref{sec:PixelVis} about Fig. \ref{fig:PixelBased}, the visual observation suggests to ``trust more the models
trained with short interval lengths.'' To an ML model-developer, this is an important hypothesis. The traditional evaluation methods based on statistical analysis cannot be used to evaluate this hypothesis easily because one would need an algorithm to find all mood change parts in the dataset to calculate some statistics, but the ``global'' ground truth labels cannot support such an algorithm.

On the other hand, it is rather easy to create a dual-flux ThemeRiver and pixel-based visualization for any music. One can quickly identify parts of music featuring noticeable mood changes using the dual-flux ThemeRiver. Fig. \ref{fig:Validation} shows 12 such examples.
One can observe the pixel-based visualization corresponding to each change to see how models cast their votes.
Same as Fig. \ref{fig:PixelBased}(d), the pixel-based visualizations in Fig. \ref{fig:Validation} are sorted by interval length first and then accuracy.

For each example, we identify which vertical parts of the pixel-based visualization correlate with the corresponding dual-flux ThemeRiver better.
For example, interval lengths $4\sim12$ for ex1, $4\sim14$ for ex9, $3\sim10$ for ex10, and  $1\sim9$ for ex12 provide positive evidence to support the hypothesis.
The visualizations for ex3, ex4, ex7, and ex11 show evidence leaning towards positive, while the visualizations for ex2, ex5, and ex6 are not indicative. 
Meanwhile, the pixel visualization for ex8 seems to provide negative evidence against the hypothesis.

Given such observation, in conjunction with other sorting strategies as shown in Fig. \ref{fig:PixelBased}, the ML model developers can make combined use of different knowledge to reason these visualizations.
It is particularly helpful when the model developers have some music knowledge or have access to such expertise to evaluate the dominant votes in the dual-flux ThemeRiver plots.

\subsection{Analytical Evaluation}
\label{sec:AnalyticalEvaluation}
%
%

Chen and Ebert proposed a systematical design and evaluation methodology \cite{chen2019ontological}, which was adopted to analyze the \emph{symptom}, \emph{cause}, \emph{remedy}, and \emph{side-effects} of different approaches in this paper. We first examine the three visual designs discussed in Sections \ref{sec:ThemeRiver} and \ref{sec:Op57} and then the combined use of pixel-based visualization and dual-flux ThemeRiver discussed in Sections \ref{sec:PixelVis} and \ref{sec:Validation}. 

\vspace{2mm}
\noindent \textbf{Three Visual Designs.}
For the requirements described in Section \ref{sec:EnsembleModels}, the three line-graph based visual design (i.e., stacked line graph, original ThemeRiver, and dual-flux ThemeRiver) cannot support R$_6$ or R$_7$ easily.
Although all three visual designs convey more or less the same amount of information, they have different strengths and weaknesses in supporting R$_{1\sim5}$.

\begin{enumerate}
    \vspace{-2.5mm}
    \item \textbf{Symptom:} With the stacked line graph and the original ThemeRiver, one cannot see easily the changes in the dominant opinion, the ordering of other opinions, and the place where ordering changes. Observing such information is an essential part of R$_{1\sim5}$.%
    \vspace{-2.5mm}
    \item \textbf{Cause:} Although such information is depicted implicitly, the cognitive cost for gaining it is very high, as it would involve perceptual estimation of the heights of different cross-sections, and cognitive comparison of such height measures \cite{Borgo:2010:TVCG}. The stacked line graph has some advantages over the original ThemeRiver in estimating the total height and that of the bottom stream.
    \vspace{-2.5mm}
    \item \textbf{Remedy:} Introduce a more explicit depiction of such information to reduce the cognitive cost. With the dual-flux ThemeRiver, the dominant opinion, the ordering, and the places of mood changes are all explicit, ready to be perceived.
    \vspace{-2.5mm}
    \item \textbf{Side-effect:} The mood streams are no longer continuous, and it may take extra effort to re-connecting the same stream e.g., to quantify the amount of mood change. With only four moods and appropriate color-coding, the side-effect is not a big issue. It could become more serious if there were many streams. 
\end{enumerate}

\noindent\textbf{Pixel-based Visualization and dual-flux ThemeRiver.} This combined use of two visual designs is for supporting requirements R$_6$ and R$_7$. To address the issue, we have to go back to the traditional methods for observing ML models' performance. When one has a few models to compare, one might be able to ensure the demanding effort for observing their performance against individual data objects (music clips in this work) by reading classification logs. However, this would not scale up to 210 ML models.

\begin{enumerate}
    \vspace{-2.5mm}
    \item \textbf{Symptom:} It is almost impossible to observe a large number of ensemble models against individual data objects by reading classification logs.
    \vspace{-2.5mm}
    \item \textbf{Cause:} It incurs very high cognitive costs of reading numbers, remembering them for building up a mental overview model, and performing comparative tasks mentally.
    \vspace{-2.5mm}
    \item \textbf{Remedy:} Both pixel-based visualization and dual-flux ThemeRiver provide external memorization, substantially reducing the cost of repeated reading-remembering. By removing the burden of memorization, the users can devote more cognitive resources to the patterns depicted.
    \vspace{-2.5mm}
    \item \textbf{Side-effect (new symptom):} Identifying individual ML models is difficult with an arbitrary list of models, and grouping models visually is even harder.
    \vspace{-2.5mm}
    \item \textbf{Cause:} Labelling small pixels is not easy. Visual grouping demands extra cognitive load for remembering and formulating groups mentally.
    \vspace{-2.5mm}
    \item \textbf{Remedy:} Using different sorting schemes.
    \vspace{-2.5mm}
    \item \textbf{Side-effect:} There could be an issue if the sorting scheme is unfamiliar to a user. For ML model-developers, this is unlikely.
\end{enumerate}

The above analytical evaluation shows that the merits of the proposed dual-flux ThemeRiver and the combined use of pixel-based visualization and dual-flux ThemeRiver can be reasoned analytically according to information theory and cognitive theories. 

\subsection{Expert Evaluation}
\label{sec:ExpertEvaluation}
To evaluate the effectiveness of our visual designs, we conducted informal interviews with 6 independent experts from music and ML domains. We met the experts individually via video conferencing. We first introduced the background of our research, then showed the video with audio for music as discussed in Section \ref{sec:Op57} and the comparisons with 12 examples as discussed in Section \ref{sec:Validation}.

Three music experts (E1, E2, E3), who major in piano, cello, and Erhu, received professional music training for over 20 years.
Three ML experts (E4, E5, E6) are specialized in different ML areas.
E4 is specialized in natural language processing (NLP) and was able to relate ML applications in music with languages. 
E5 is an expert in convolutional neural networks and is aware of ensemble modelling and the benefits of VIS4ML techniques.
E6 is an expert in ML and optimization and is mathematically knowledgeable.
We summarize their main points from three perspectives, music, visualization, and ML. Each point is labelled with a circled letter (e.g., \textcircled{a}), transcribed comments are in italic, added clarifications are in [], and the authors' feedback and actions are marked with $\blacktriangleright$.  


\vspace{2mm}
\noindent\textbf{From the Music Perspective:}

\textcircled{a} E1 considered that \emph{the dual-flux ThemeRiver avoids confusion. The original ThemeRiver can make people think that the bottom color is the base mood or the main mood, and the top mood is minor.}%

\textcircled{b} E2 noted that the dual-flux ThemeRiver can be applied to a \textbf{music ensemble} with multiple instruments \emph{so that we can see how the individual instrument produces the mood and how the mood is created when multiple instruments are played simultaneously. It will be interesting for music analysis.}

\textcircled{c} Five experts were in favor of keeping the minor mood, even if a mood attracts only a few votes. \emph{We need to have some flexibility. And the music itself is subjective.} \emph{I have seen people from other cultures considering the major key to be sad and the minor key to be happy. So if we were to use their labeling data for training, the results would be the opposite.} The only exception is E2, who considered scenarios where \emph{some music is designed to be sad or has a specific mood. ...
So these opinions are noise and are okay to be ignored.}

    $\blacktriangleright$ There is a general debate about the principle that different moods can co-exist. We concluded that E2's specific scenarios should not restrict the flexibility favored by others. In a scenario where only one mood is allowed per time step, the musicians usually have the knowledge to ignore the other opinions. This can be done easily by focusing on the dominant stream at the top half of a dual-flux ThemeRiver.    

\textcircled{d} E3 noticed the fact that the labelled data has only 4 classes, and would like to see more musical words, such as \emph{Affettuoso (excited), Con moto (vivid), and Con tenerezza (gentle).}

    $\blacktriangleright$ There are currently no labelled data for the over 60 music words mentioned by E3. Nevertheless, this will be an important future direction for all of us working on music-related ML.

\vspace{2mm}
\noindent\textbf{From the Visualization Perspective:}

\textcircled{e} In general, all experts agreed that the dual-flux ThemeRiver can depict the flow of mood changes while conveying the uncertainty of mood classification.

\textcircled{f} Five experts considered that the dual-flux ThemeRiver helped determine the dominant mood more easily. The only exception was E1 who considered that it was difficult to learn.
E2 and E3 commented on the learning aspect: \textit{As we are not professional visual analysts, we need to be trained to read the dual-flux ThemeRiver as it seems a bit more complicated with lines and switches. Once we know how to read it, it is much easier to identify the dominant mood and the uncertainty.}
 
 \textcircled{g} E3 suggested a new variant of the dual-flux ThemeRiver. \textit{"Maybe we can have ... one that has the same time scale as the music score."}.

    $\blacktriangleright$ We researched this suggestion.
    In most visual representations of sound, the temporal length is encoded as spatial length, while in a music score, the temporal length of a music note is encoded symbolically (e.g., semibreve, minim, crotchet, quaver). If one wishes to synchronize both representations, one could distort either the spatial lengths of the time steps in a sound representation or the spatial gaps between music notes in a music score.  
    Perceptually, the relative merits might be in favor of the latter. We believe that further cognitive research will be needed before releasing E3's variant as a new music representation. 

\textcircled{h} E4 and E5 suggested that to reduce the number of mood switches by \emph{setting a threshold of maybe 5\% before letting it switch.}

    $\blacktriangleright$ We investigated this suggestion carefully.
    We consider that such a threshold would undermine the interpretation of the dominant mood as well as the ordering of the moods. The threshold will likely be a hidden parameter, and there may need an additional parameter for determining how long a switching may be delayed.    
    We went back to E4 and E5 with our investigation. They both agreed that this could lead to new problems, confusion, and extra learning cost.

\vspace{2mm}
\noindent\textbf{From the ML Perspective:}

\textcircled{i} ML experts (E4, E5, and E6) are positive about the dual-flux ThemeRiver for conveying ML predictions with uncertainty and its combination with the pixel-based visualization, as the combined design enables ML developers to extract information more efficiently for understanding and \emph{optimization of ensemble learning}.

\textcircled{j} There were two opinions as to whether less accurate models should be eliminated. E6 found that it was easy to observe poor performance and reason about the causes using the pixel-based visualization, and was in support of elimination. E4 argued for keeping all the models as part of the ensemble learning.

\textcircled{k} Both E4 and E6 agree on moderating using weights as votes. E4 commented: \emph{It is okay to adjust the weights, but we should keep them as it is already the trained results.} E6 commented: \emph{We may set some conditions to adjust the weights.}

    $\blacktriangleright$ We adopted the suggestion and extended the uses of pixel-based visualization for showing the weights as detailed below.

\begin{figure}[t]
    \begin{tabular}{@{}r@{\hspace{5mm}}r@{\hspace{1mm}}r@{\hspace{1mm}}l@{\hspace{1mm}}r@{\hspace{1mm}}l@{\hspace{1mm}}r@{\hspace{1mm}}l@{}}
        \includegraphics[width=15mm,height=6mm]{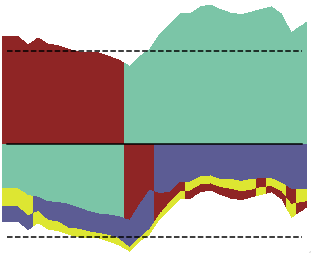} & &
        \includegraphics[width=15mm,height=6mm]{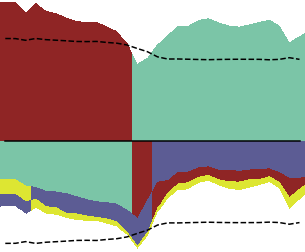} & &
        \includegraphics[width=15mm,height=6mm]{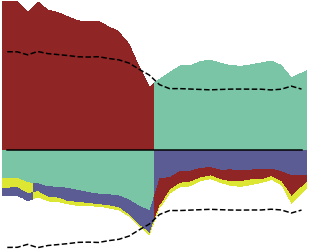} & &
        \includegraphics[width=15mm,height=6mm]{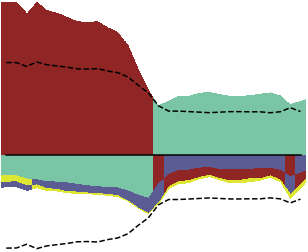} &\\
        \includegraphics[width=16mm, height=65mm]{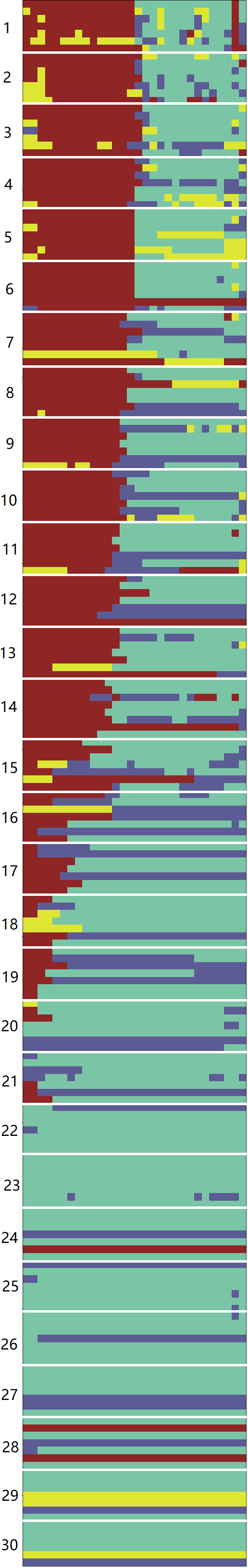} & 
        \includegraphics[width=2mm,height=65mm]{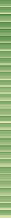} & \includegraphics[width=15mm,height=65mm]{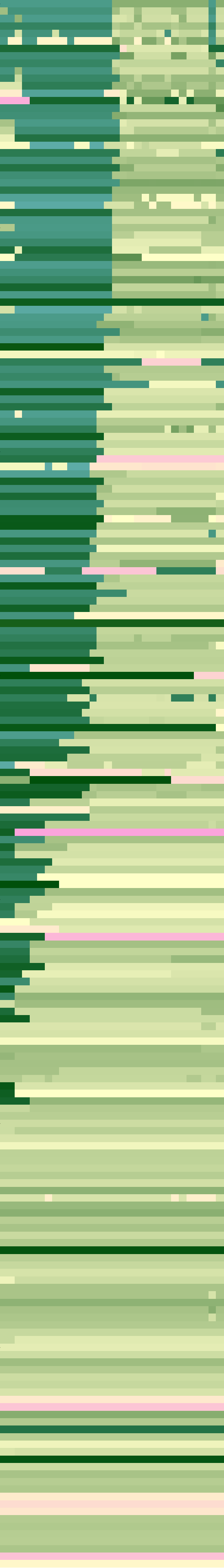} &
        \includegraphics[width=2mm,height=65mm]{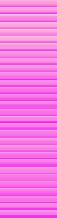} & \includegraphics[width=15mm,height=65mm]{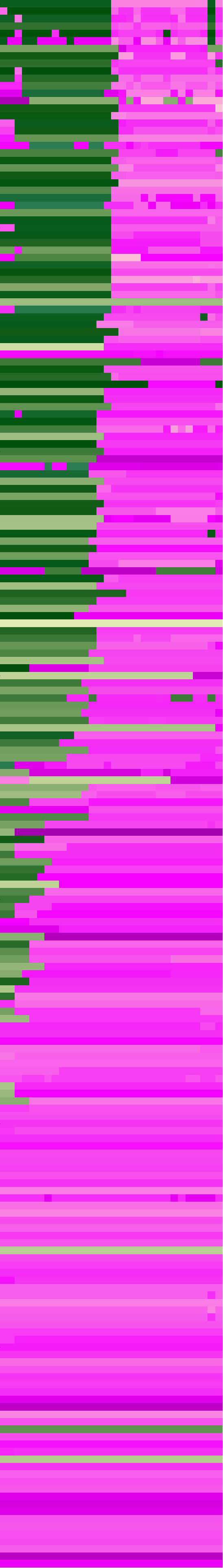} &
        \includegraphics[width=2mm,height=65mm]{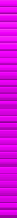} & \includegraphics[width=15mm,height=65mm]{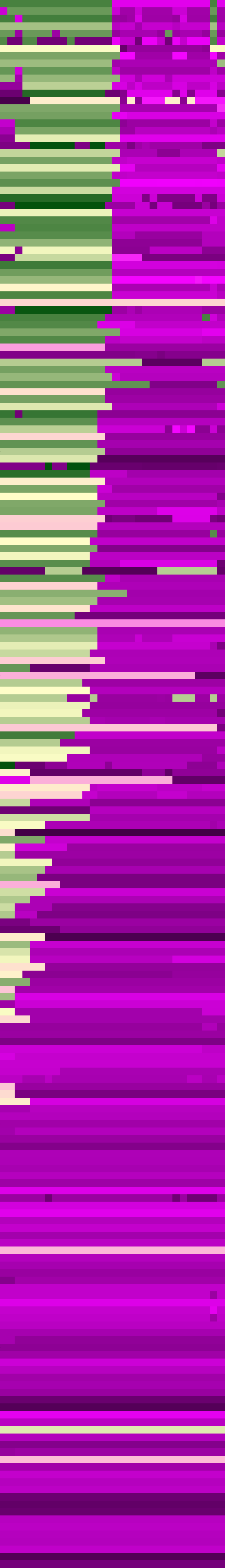}
    \end{tabular}\\
    \centering
        \includegraphics[height=3mm]{Figures/legend_row.png}\\
        \includegraphics[height=3.5mm]{Figures/legend_color.png}
    \centering
    \caption{The first column is from Fig \ref{fig:PixelBased}(d). On its right, the three wide pixel-maps show the weighted class-accuracy $\alpha$ of each prediction (i.e., a pixel in the first column). The three weighting functions are $\alpha$, $\alpha^2$, and $\alpha^3$ respectively. They were used to compute the weighted dual-flux ThemeRiver above these wide pixel-maps. On the left of each pixel-map, a narrow pixel-bar shows the weighted model accuracy.}
    \label{expetreviewexample}
    \vspace{-4mm}
\end{figure}




As described in sections \ref{sec:ThemeRiver} and \ref{sec:PixelVis}, dual-flux ThemeRiver can handle weighted votes. As shown in Fig. \ref{fig:PixelBased}, one approach is to assign the weight to each model decision based upon the class-accuracy of the model.
As shown in Fig. \ref{expetreviewexample}, we can also use pixel-maps to compare different weighting schemes. For example, let $\alpha_{i,j}$ be the class-accuracy associated with each pixel. The first wide pixel-map on the left in Fig. \ref{expetreviewexample} shows the prediction of each pixel. The second, third, and Fourth wide pixel-maps show the values of $\alpha_{i,j}$, $\alpha^2_{i,j}$, and $\alpha^3_{i,j}$ respectively. Between these pixel-maps are the narrow pixel-bars for showing the overall accuracy of each model, and its squared and cubed versions in the second and third narrow pixel-bars.
The dual-flux ThemeRiver plots on the top show the effects of different weighting schemes, i.e., from left: unweighted, $\alpha_{i,j}$, $\alpha^2_{i,j}$, and $\alpha^3_{i,j}$. 
Experts can evaluate different weighting schemes and identify a suitable one for describing the mood.

We showed this result to E3 and E6. Both quickly noticed the reduction of calm in the second half of the music. They commented: \textit{We understand the increase in ``angry''}. \emph{We might say that it is a ``stronger angry'' mood, but not ``stronger calm''}.
E3 commented further: \emph{I would say the weights of squared [class-]accuracy fit better. I assume that the calm mood itself is weaker, so when it is calm, it should be kept at a moderate level as ``standard'' instead of the plots shown with the other two weighting schemes. I am happy with the flexibility to have different options. For now, the squared [class-]accuracy should fit the best."}

\section{Conclusions}
\label{sec:Conclusions}

In this work, we have developed visualization solutions for supporting the development and analysis of ensemble ML models for music mood classification as well as for communication of ensemble model predictions.
%
To address the challenge of training models with less accurate ``global'' labels, we appreciate the need for users to make sense of the predictions by a large collection of ML models, for which VIS is essential.
As shown in Section \ref{sec:Op57}, those with music knowledge can reason the ensemble ML predictions using the dual-flux ThemeRiver plot, which therefore provides a means for enabling expressive AI or the understanding of AI. 

As exemplified in Sections \ref{sec:PixelVis} and \ref{sec:Validation}, when ML model-developers examine the behaviours of one or a few ML models, 
they will frequently formulate hypotheses about the models (e.g., poorly-performed training methods, unsuitable time intervals, and optimal weighting function).
One major advantage of using VIS is that the ML model-developers can quickly inspect the behaviours of many models. Such visual evaluation of a hypothesis is highly cost-effective if the VIS techniques are readily available, despite it may not be the final step for confirming or falsifying the hypothesis concerned.

Any slightly complex visualization incurs some information loss in comparison with the original data \cite{Chen:2016:TVCG}. Inevitably it will have some limitations for some tasks. The analytical and expert evaluation in Sections \ref{sec:AnalyticalEvaluation} and \ref{sec:ExpertEvaluation} enabled us to identify such limitations and side-effects. In some cases, we have improved our work. In other cases, the benefits out-weighed the limitations.

We are in the process of using the VIS techniques developed in this work as well as others in the literature to study a much larger ensemble of ML models for music mood classification, including models developed using other ML frameworks. We also anticipate the potential use of the dual-flux ThemeRiver design in other applications, such as visualizing the results of opinion polling and the emotion featured in texts and videos (e.g., \cite{xia2020seqdynamics, zeng2020emotioncues}). In such visualization, it is important to observe the ordering of different opinions or emotion streams. We hope that future work will explore such opportunities.

\newpage



\section*{Acknowledgement}
Part of this work was made possible by the Network of European Data Scientists (NeEDS), a Research and Innovation Staff Exchange (RISE) project under the Marie Skłodowska-Curie Program.

\bibliographystyle{abbrv-doi}

\bibliography{refs}
\end{document}